\documentclass{article}
\usepackage{arxiv}

\usepackage[utf8]{inputenc} 
\usepackage[T1]{fontenc}    
\usepackage{hyperref}       
\usepackage{url}            
\usepackage{booktabs}       
\usepackage{amsfonts}       
\usepackage{nicefrac}       
\usepackage{microtype}      
\usepackage{lipsum}		
\usepackage{graphicx}
\usepackage{natbib}
\usepackage{doi}
\usepackage{amsmath}
\usepackage{amssymb}
\usepackage{enumitem}
\usepackage{pdfpages}
\usepackage{blkarray}
\usepackage{mathtools}
\usepackage{natbib}
\usepackage{url}
\usepackage{graphicx,subfigure}
\usepackage{epstopdf} 
\usepackage{array}

\def\bSig\mathbf{\Sigma}

\usepackage{lipsum}
\usepackage{setspace}
\usepackage{xcolor}
\usepackage{url}
\usepackage{hyperref}
\usepackage{bm}
\pdfoutput=1
\usepackage{amsmath,amssymb, amsfonts}
\usepackage{algorithm}
\usepackage{algpseudocode}
\usepackage[format=plain]{caption}
\usepackage{animate}
\usepackage{tikz}
\usepackage{amsmath}
\usepackage{graphicx}
\usepackage{natbib}
\usepackage{graphics}

\newcommand\highlight[3]{\colorbox{#1}{\textcolor{#2}{#3}}}

\def\bmu{\mbox{\boldmath{$\mu$}}}

\newcommand{\bGamma}{\mbox{\boldmath{${\Gamma}$}}}
\newcommand{\bbeta}{\mbox{\boldmath{$\beta$}}}

\newcommand{\bgamma}{\mbox{\boldmath{${\gamma}$}}}

\def\smt{{\mbox{\tiny T}}}

\def\bX{\mathbf X}
\def\bx{{\bf x}}

\def\bZ{{\bf Z}}

\def\bSb{\mathbf{S}_{b}}
\def\bSw{\mathbf{S}_{w}}
	
\def\bSp{\mathbf{S}_{p}}	
\def\bphi{\mbox{\boldmath{$\phi$}}}
\def\bPhi{\mbox{\boldmath{$\Phi$}}}
\def\bA{\mathbf{A}}

\def\bA{\mathbf A}
\def\bB{\mathbf B}

\def\bX{\mathbf X}
\def\bx{\mathbf x}

\def\bS{\mathbf S}

\def\hbmu{\mbox{\boldmath{$\hat{\bmu}$}}}

\usepackage{color}
\usepackage{comment}
\usepackage{ulem}
\usepackage{soul}

\definecolor{gw}{RGB}{10,101,181}

\graphicspath{{images/}}

\DeclareUnicodeCharacter{2032}{XXXXXX}

\title{Multivariate Functional Linear Discriminant Analysis with Feature Selection: An Application to Inflammatory Bowel Disease Classification}

\author{{\hspace{1mm}Limeng Liu} \\
	Division of Biostatistics and Health Data Science\\
	University of Minnesota Twin Cities\\
	Minneapolis, MN 55455 \\
	\texttt{liu02356@umn.edu} \\
	\And
	{\hspace{1mm}Guannan Wang} \\
	Department of Mathematics\\
	William \& Mary\\
	Williamsburg, VA 23185 \\
	\texttt{gwang01@wm.edu} \\
        \And
	{\hspace{1mm}Sandra E. Safo}\thanks{Corresponding Author: Sandra E. Safo, www.sandraesafo.com} \\
	Division of Biostatistics and Health Data Science\\
	University of Minnesota Twin Cities\\
	Minneapolis, MN 55455 \\
	\texttt{ssafo@umn.edu} \\
}

\renewcommand{\shorttitle}{Multivariate Functional Linear Discriminant Analysis}

\begin{document}
\maketitle

\begin{abstract}
Inflammatory Bowel Disease (IBD), including Crohn's Disease (CD) and Ulcerative Colitis (UC), presents significant public health challenges due to its complex etiology. Motivated by the IBD study of the Integrative Human Microbiome Project, our objective is to identify microbial pathways that distinguish between CD, UC, and non-IBD over time. Most current research relies on simplistic analyses that examine one variable or time point at a time, or address binary classification problems, limiting our understanding of the dynamic interactions within the microbiome over time. To address these limitations, we develop a novel functional data analysis approach for discriminant analysis of multivariate functional data that can effectively handle multiple high-dimensional predictors, sparse time points, and categorical outcomes. Our method seeks linear combinations of functions (i.e., discriminant functions) that maximize separation between two or more groups over time. We impose a sparsity-inducing penalty when estimating the discriminant functions, allowing us to identify relevant discriminating variables over time. Applications of our method to the motivating data identified microbial features related to mucin degradation, amino acid metabolism, and peptidoglycan recognition, which are implicated in the progression and development of IBD. Furthermore, our method highlighted the role of multiple vitamin B deficiencies in the context of IBD. By moving beyond traditional analytical frameworks, our innovative approach holds the potential for uncovering clinically meaningful discoveries in IBD research.
\end{abstract} 

\section{Introduction}
\label{sec:intro}

Inflammatory Bowel Disease (IBD), which includes Crohn's Disease (CD) and Ulcerative Colitis (UC), is a complex multifactorial autoimmune disorder that affects millions of people worldwide \citep{huttenhower2014inflammatory} and presents a substantial economic and social burden \citep{malmborg2016emerging, ananthakrishnan2023economic}. IBD is characterized by chronic inflammation of the gastrointestinal tract, driven by a combination of genetic predisposition, environmental factors, immune system dysfunction, and notably, disruptions in the gut microbiome \citep{baumgart2007inflammatory, lennard1989classification}. Even though IBD is one of the most intensively studied conditions in terms of microbial-immune system imbalances \citep{hold2014role, sokol2006specificities}, the exact cause of IBD remains unknown. The interaction between the host and the microbiome, combined with the diversity and variability of microbial communities, adds considerable complexity to understanding IBD pathogenesis. This complexity makes it challenging to identify specific triggers or predict disease progression.

Many efforts have been made to better understand the role of the microbiome in various diseases, including IBD. In particular, the Integrative Human Microbiome Project (iHMP) IBD Study \citep{lloyd2019multi} was launched to investigate factors contributing to the heterogeneity in CD, UC, and non-IBD cases. The study enrolled 130 participants, both with and without IBD, from five medical centers and followed them over the course of a year (50 weeks). Each week, a comprehensive set of data was collected from patient samples—including genomic, transcriptomic, and proteomic profiles—as well as microbiome data such as metagenomic, metatranscriptomic, and metaproteomic information. This rich, longitudinal dataset offers valuable insights into functional dysbiosis of the gut microbiome associated with the onset and progression of IBD. Since IBD is a chronic condition with relatively stable diagnostic classifications over short time intervals, biological conversion from UC to CD, or vice versa, is extremely rare \citep{moss2008often, lee2016change}. Therefore, this study assumes that each patient’s diagnosis remains unchanged throughout the 50-week period. However, some participants missed scheduled assessments or were lost to follow-up, resulting in sparse and irregularly spaced observation time points. 

Other studies have also analyzed data from the iHMP IBD study and have improved our understanding of microbial features that discriminate between those with and without IBD \citep{chen2020gut, liang2021altered}. Many of these studies ignored the temporal dynamics and high dimensionality of the microbiome data and focused on an individual microbial feature or a single time point \citep{wang2024b3gnt7, liang2021altered}. Other studies have primarily focused on identifying microbial features that distinguish between individuals with and without IBD \citep{yu2023development}. However, the critical challenge of differentiating among all three classes -- UC, CD, and non-IBD -- remains underexplored. This gap is clinically meaningful, as CD and UC share many overlapping symptoms but differ significantly in disease localization: CD can affect any part of the gastrointestinal tract, whereas UC is restricted to the colon \citep{granlund2013whole}. This distinction highlights the importance of identifying robust biomarkers capable of distinguishing among CD, UC, and non-IBD cases. Addressing this challenge requires moving beyond conventional, simplistic methods and adopting advanced analytical approaches that can capture temporal correlations, accommodate multi-class classification problems, handle sparse and irregular observation time points, and scale to high-dimensional data (e.g., the IBD dataset includes 2,261 microbial pathways), while maintaining interpretability by identifying signal variables in the presence of noise.

Functional data analysis (FDA) methods have been developed to address several of these challenges. However, to the best of our knowledge, no existing method fully encompasses all these limitations simultaneously. For instance, Functional Principal Components Analysis (FPCA) \citep{rice1991estimating}, a widely used tool for analyzing functional data, can capture temporal correlations and handle sparse and irregularly sampled data (where each subject is observed at only a few irregular time points). However, it is limited in its ability to handle high-dimensional functional data and perform effective discrimination. Multivariate FPCA (MFPCA) \citep{yao2005functional, chiou2014multivariate} has been developed for multivariate functional data, addressing the high-dimensional limitations of FPCA. Because the goal of MFPCA is not discrimination, the functional principal components it produces may not effectively distinguish between groups. Functional linear discriminant analysis (FLDA) \citep{gardner2021linear} extends traditional linear discriminant analysis (LDA) to functional data and uses linear combinations of functions to discriminate between two groups over time. FLDA can model temporal correlations and is suitable for high-dimensional functional data. However, it is limited to binary classification and cannot handle data with sparse and irregular time points. A key limitation shared by these FDA methods is their limited ability to perform variable selection for group discrimination over time, which hinders interpretability and reduces the capacity to identify meaningful, clinically relevant features. 

Other FDA methods for variable selection typically focus on selecting time intervals for a single functional variable (i.e., $p = 1$) and are largely restricted to binary classification problems (e.g., \citep{park2022sparse}). Meanwhile, LDA and sparse LDA methods can perform variable selection (e.g., \cite{safo2019sparse, witten2011penalized, wen2018robust}), offer variable selection capabilities but are designed for cross-sectional data and are unable to account for temporal dependencies inherent in longitudinal functional datasets.

To achieve our goal of identifying microbial features that distinguish between UC, CD, and non-IBD over time, we develop a novel multivariate FLDA (MFLDA) method that addresses the challenges in existing methods. \textit{First,} MFLDA captures temporal correlations by treating functions as predictors (such as curves of IBD microbial pathways over time) rather than individual data points, to identify linear combinations of functions that optimally distinguish between groups.
\textit{Second,} MFLDA simultaneously considers multiple features over time, unlike single-variable models that analyze features independently, providing a more comprehensive understanding of microbiome dynamics and improving classification accuracy as the number of observed time points increases. 
\textit{Third}, MFLDA enables the integration of feature selection through an optimization problem with a sparsity-inducing penalty, enabling the identification of relevant functional relationships and enhancing interpretability. 
\textit{Fourth}, MFLDA supports multi-class classification, addressing the limitations of existing methods that are often restricted to binary outcomes. 
\textit{Finally}, MFLDA models irregularly sampled data by incorporating spline smoothing techniques. MFLDA extends existing FDA approaches to effectively handle and classify irregularly sampled data, which are common in real-world longitudinal datasets.
Our simulations show that MFLDA is effective at multi-class functional discrimination, outperforming other existing methods. When we applied MFLDA to the microbiome data from the iHMP Study, we identified pathways -- including those related to multiple vitamin B deficiencies -- that distinguish between UC, CD, and non-IBD.

The remainder of the paper is organized as follows. In Section \ref{sec:meth}, we introduce the proposed methods. In Section \ref{sec:simulation}, we conduct simulation studies to assess the performance of our methods in comparison with other methods. In Section \ref{sec:IBD}, we apply our method to the motivating data, and we provide concluding remarks in Section \ref{sec:conclusion}.

\section{Methodology}
\label{sec:meth}

\subsection{General Notations and Goal}
In this subsection, we introduce the general notation for our proposed method. Denote the high-dimensional multivariate longitudinal predictors (e.g., microbiome data) at time $t$ by $\bX(t) = (\bx_1(t), \cdots, \bx_n(t))^{\smt} \in \Re^{n \times p}$, where $\bx_i(t) \in \Re^p$, $i = 1, \ldots, n$. There are $p$ microbial pathways measured on $n$ individuals at time $t \in [0, T]$. In our application, the maximum number of time points is $T=50$ weeks, with $p = 2,261$ microbial pathways and $n = 130$ individuals. As noted, our method does not require regular or identical time points for individuals. For simplicity, we use $t$ to represent $t_i$, $i = 1, \ldots, n$. For each individual $i$, let $y_i(t)$ denote their class membership $k$, $k = 1, \ldots, G$, at time $t$. For example, $G = 2$ for IBD and non-IBD, or $G = 3$ for CD, UC, and non-IBD. In our motivating data, $y_i(t) = y_i$, indicating that class membership does not change over time for individual $i$. In other contexts, class membership may vary over time, and our method can accommodate time-dependent class membership as long as there are $G$ groups at each time point. Given data on class membership and longitudinal predictors, we aim to: i) predict the class membership $y_j$ of a new subject $j$ using their high-dimensional predictors, and ii) determine relevant microbial pathways that discriminate by IBD status over time to better understand its pathobiology.

\subsection{Multivariate Linear Discriminant Analysis}

To introduce our methodology, we first assume $t = 1$, reducing the data to a cross-sectional setting (i.e., $\bX(t) = \bX$). In this subsection, we begin by introducing the traditional multivariate LDA method, which focuses on cross-sectional data. In subsequent sections, we will extend these techniques to the FDA framework. Let $\bX = (\bX_{n_1}, \ldots, \bX_{n_k}, \ldots, \bX_{n_G})$ be the concatenation of data from all classes, where ${\bX_{n_k}} = (\bx_{1k}, \ldots, \bx_{n_k})^{\smt} \in \Re^{n_k \times p}$. Here, $\bx_{ik} \in \Re^p$ is the data vector for the individual $i$ in class $k$, $k = 1, \ldots, G$, $G_k$ is the set of samples that belongs to the class $k$ and $n_k$ the number of samples in class $k$. Then, the mean vector for class $k$, the common covariance matrix for all classes, and the covariance between classes are given, respectively, by $\hat{\bmu}_k = (1/n_k)\sum_{i \in G_k}\bx_{ik}$; $\bSw = \sum\limits_{k=1}^G \sum\limits_{i \in G_k}(\bx_{ik}-\hat{\bmu}_k)(\bx_{ik}-\hat{\bmu}_k)^{\smt}$; $\bSb = \sum\limits_{k=1}^G n_k(\hat{\bmu}_k-\hat{\bmu})(\hat{\bmu}_k-\hat{\bmu})^{\smt}$. Here, $\hat{\bmu}$ is the combined class mean vector and is defined as $\hbmu = (1/n)\sum\limits_{k=1}^{G} n_k\hbmu_k$. For a $G$ class prediction problem, LDA finds $G-1$ discriminant vectors $\bB = [\bbeta_1, \ldots, \bbeta_{G-1}] \in \Re^{p \times (G-1)}$, which are linear combinations of all available features such that the projected data $\bZ = \bX \bB \in \Re^{n \times (G-1)}$, also known as discriminant scores, have maximum separation between classes and minimal separation within classes. Mathematically, the discriminant vectors are the solutions to the optimization problem: $\max_{\bbeta_k} \bbeta_k^{\smt} \bSb \bbeta_k~\mbox{subject to}~ \bbeta_k^{\smt} \bSw \bbeta_k = 1, ~\bbeta_l^{\smt} \bSw \bbeta_k = 0~\forall l < k, ~k = 1, 2, \ldots, G-1$. One can show, using Langragian multipliers, that the first LDA vector can be obtained from solving the generalized eigenvalue-eigenvector system: $\bS_b \bbeta_1 = \lambda_1 \bS_w \bbeta_1$ for $(\widehat{\lambda}_1, \widehat{\bbeta}_1)$, which are the first eigenvalue-eigenvector pair of $\bSw^{-1}\bSb$ for $\bSw \succ 0$. Subsequent eigenvectors could be obtained such that $\bbeta_l^{\smt} \bS_w \bbeta_k = 0, l \ne k$. 

\subsection{Multivariate Functional Linear Discriminant Analysis}

Building on the traditional multivariate LDA method for cross-sectional data \citep{SafoBiomSELP}, our approach extends it to the multivariate functional data setting. In addition to accommodating functional data, our method incorporates feature selection and enhances interpretability, addressing key limitations of the classical approach. For individual $i$, $i = 1, \ldots, n$ with $j = 1, \ldots, p$ features measured at time $t$, denote their observed data by $x_{ij}(t)$. We assume that there exists a continuous mean function $w_{ij}(t)$ such that the observed data are the sum of this function and measurement error $\epsilon_{ij}(t)$, i.e., $x_{ij}(t) = w_{ij}(t) + \epsilon_{ij}(t)$. The mean function can be estimated using the spline smoothing method described in \citep{wang2015spline}. Spline methods are widely used nonparametric techniques, offering flexible and accurate approximations of smooth functions by representing them as linear combinations of spline basis functions. Then, $w_{ij}(t)$ can be approximated as $w_{ij}(t) = [\phi_1(t), \ldots, \phi_m(t)] \mathbf{c}_{ij} = \bphi^{\smt}(t) \mathbf{c}_{ij}$, where $(\phi_1(t), \ldots, \phi_m(t))$ is the set of $m$ basis functions. The number of $m$ is chosen based on type of spline and number of interior knots. The spline coefficient matrix $\mathbf{C}_i = [\mathbf{c}_{i1}, \mathbf{c}_{i2}, \ldots, \mathbf{c}_{ip}] \in \Re^{m \times p}$ can be estimated with least squares or maximum likelihood method. Here, we estimate the coefficient matrix by least squares \citep{wang2015spline}. When time points are sparse or irregular, estimation can be performed by pooling information across subjects, provided that each subject has a minimum number of observations (or minimum number of time points) determined by spline parameters (i.e. minimum number of time points should be greater than $m$).

Multivariate LDA seeks linear combinations (weights) of $p$ features with maximum separation between classes and minimum separation within classes. These linear combinations are assumed to be independent of each other, for uniqueness. In the functional setting, for the first discriminant function, we seek \textbf{linear combinations of $p$ functions that can discriminate between functions in different groups}. Denote the first functional discriminant vector at time $t$ by $\bbeta(t)$. Denote the functional discriminant score for subject $i$ at time $t$ by $z_i(t)$. Then $z_i(t) = \beta_1(t)w_{i1}(t) + \beta_2(t)w_{i2}(t) + \ldots + \beta_p(t)w_{ip}(t) = [w_{i1}(t) \ldots w_{ip}(t)] \bbeta(t) = [\bphi^{\smt}(t)\mathbf{c}_{i1}, \ldots, \bphi^{\smt}(t)\mathbf{c}_{ip}]\bbeta(t)$. We write the between-class separation in terms of $\bphi(t)$ and $\mathbf{C}_i$. To make $\bphi^{\smt}(t)\mathbf{c}_{ip}$ more specific to each class, denote it as $\bphi^{\smt}(t)\mathbf{c}_{ip}^{(k)}$ for class $k$. Thus, $z_i^{(k)}(t) = [\bphi^{\smt}(t)\mathbf{c}_{i1}^{(k)} \ldots \bphi^{\smt}(t)\mathbf{c}_{ip}^{(k)}]\bbeta(t)$. Let the class $k$ basis vector for the feature $j$ be $\mathbf{\Bar{c}}_j^{(k)} = \frac{1}{n_k} \sum_{i \in G_k} \mathbf{c}_{ij}$.
Thus, the mean functions for class $k$ can be written as $\Bar{z}_k(t) = \frac{1}{n_k} \sum_{i \in G_k} [\bphi^{\smt}(t)\mathbf{c}_{i1}^{(k)} \ldots \bphi^{\smt}(t)\mathbf{c}_{ip}^{(k)}] \bbeta(t) = [\bphi^{\smt}(t)\mathbf{\Bar{c}}_1^{(k)} \ldots \bphi^{\smt}(t)\mathbf{\Bar{c}}_p^{(k)}]\bbeta(t)$ and the overall mean function is $\Bar{z}(t) = \sum_{k=1}^G n_k \Bar{z}_k(t) / \sum_{k=1}^G n_k = [\bphi^{\smt}(t)\mathbf{\Bar{c}}_1 \ldots\bphi^{\smt}(t)\mathbf{\Bar{c}}_p]\bbeta(t)$. In particular, the separation between classes for the $k$th class is 

{\setstretch{1.0}
\begin{align*}
    \Bigl(\Bar{z}_k(t) - \Bar{z}(t)\Bigr)^{\smt}\Bigl(\Bar{z}_k(t) - \Bar{z}(t)\Bigr) &= \bbeta^{\smt}(t)
    \begin{bmatrix} 
    (\bar{\mathbf{c}}_1^{(k)} - \bar{\mathbf{c}}_1)^{\smt} \bphi (t) \\
    \vdots \\
    (\bar{\mathbf{c}}_p^{(k)} - \bar{\mathbf{c}}_p)^{\smt} \bphi (t) 
\end{bmatrix} 
\left[\bphi^{\smt}(t)(\bar{\mathbf{c}}_1^{(k)} - \bar{\mathbf{c}}_1) \ldots \bphi^{\smt}(t)(\bar{\mathbf{c}}_p^{(k)} - \bar{\mathbf{c}}_p)\right] \bbeta(t) \\
&= \bbeta^{\smt}(t) \bA_k(t) \bbeta(t),
\end{align*}
where the between-class covariance for class $k$, $\mathbf{A}_k(t) \in \Re^{p \times p}$, is given by 
$$\mathbf{A}_k(t) = \begin{bmatrix} 
    (\mathbf{\bar{c}}_1^{(k)} - \mathbf{\bar{c}}_1)^{\smt}\\
    \vdots \\
    (\mathbf{\bar{c}}_p^{(k)} - \mathbf{\bar{c}}_p)^{\smt}
\end{bmatrix} 
\bPhi(t) \bPhi(t)^{\smt}
\left[(\mathbf{\bar{c}}_1^{(k)} - \mathbf{\bar{c}}_1) \ldots (\mathbf{\bar{c}}_p^{(k)} - \mathbf{\bar{c}}_p)\right].
$$
For time points $t \in [0, T]$, let 
$\bPhi^{\smt} = \begin{bmatrix} 
    \phi_1(t_1) & \dots & \phi_m(t_1) \\
    \vdots & \ddots & \vdots\\
    \phi_1(t_T) & \dots & \phi_m(t_T)
    \end{bmatrix} \in \Re^{T \times m}$ 
be a matrix} that collects the $m$ basis functions in $T$ time points. 

Thus, the overall between-class covariance matrix, $\bSb \in \Re^{pT \times pT}$,  is defined as $\bSb = \sum_{k = 1}^G n_k \bA_k(t).$
We then shift our focus to the within class variance. The \textbf{variance covariance function} between two features $j$ and $q$ for time point $t_h$, $t_s$ is $s_{jq}(t_h,t_s)$. This holds for all pairs \(j, q = 1, \dots, p\) and all pairs \(t_h, t_s = 1, \dots, T\). We construct the within-class covariance matrix for class $k$, i.e., $\bS^{(k)} \in \Re^{pT \times pT}$. Let $\bS^{(k)}(t_h, t_s) \in \Re^{p \times p}$ be the $t_ht_s$-entry of $\bS^{(k)}$. The mean of \(w_{ij}(t_h)\) in group \(k\) is given by $\bar{w}_{kj}(t_h) = \frac{1}{n_k} \sum_{i \in G_k} w_{ij}(t_h)$, and the mean of \(w_{ij}(t_s)\) in group \(k\) is given by $\bar{w}_{kj}(t_s) = \frac{1}{n_k} \sum_{i \in G_k} w_{ij}(t_s)$. The $jq$th entry in $\bS^{(k)}(t_h,t_s)$ is: 

{\setstretch{1.0}$$\begin{aligned}
s_{jq}^{(k)}(t_h,t_s) &= \frac{1}{n_k - 1} \sum_{i \in G_k} \left(w_{ij}(t_h) - \frac{1}{n_k} \sum_{i \in G_k} w_{ij}(t_h)\right)^{\smt} \left(w_{iq}(t_s) - \frac{1}{n_k} \sum_{i \in G_k} w_{iq}(t_s)\right) \\
&= \frac{1}{n_k - 1} \sum_{i \in G_k} \Bigl(w_{ij}(t_h) -\bar{w}_{kj}(t_h)\Bigr)^{\smt} \Bigl(w_{iq}(t_s)-\bar{w}_{kq}(t_s)\Bigr).
\end{aligned}$$}

Since $w_{ij}(t_h)$ can also be written as $\bphi^{\smt}(t_h)c_{ij}^{(k)}$ for each group $k \in G_k$, the mean in each group $k$ is $\bar{w}_{kj}(t_h) = \frac{1}{n_k} \sum_{i \in G_k} \bphi^{\smt}(t_h)c_{ij}^{(k)} = \bphi^{\smt}(t_h)\bar{\mathbf{c}}_j^{(k)}$. Similarly for $w_{ij}(t_s)$ and $\bar{w}_{kj}(t_h)$. Then the above can be further written as

{\setstretch{1.0}$$\begin{aligned}
s_{jq}^{(k)}(t_h, t_s) &= \frac{1}{n_k - 1} \sum_{i \in G_k} \left(\bphi^{\smt}(t_h)c_{ij}^{(k)} -\bphi^{\smt}(t_h)\bar{\mathbf{c}}_j^{(k)}\right)^{\smt} \left(\bphi^{\smt}(t_s)c_{iq}^{(k)} -\bphi^{\smt}(t_s)\bar{\mathbf{c}}_q^{(k)}\right)\\
&= \frac{1}{n_k - 1} \sum_{i \in G_k}
\Bigl(c_{ij}^{(k)} - \bar{\mathbf{c}}_j^{(k)}\Bigr)^{\smt} \bphi(t_h) \bphi(t_s)^{\smt} \Bigl(c_{iq}^{(k)} - \bar{\mathbf{c}}_q^{(k)}\Bigr).
\end{aligned}$$}

When $j = q$, $s_{jj}^{(k)}(t_h, t_s) = \frac{1}{n_k - 1} \sum_{i \in G_k}
\Bigl(c_{ij}^{(k)} - \bar{\mathbf{c}}_j^{(k)}\Bigr)^{\smt} \bphi(t_h) \bphi(t_s)^{\smt} \Bigl(c_{ij}^{(k)} - \bar{\mathbf{c}}_j^{(k)}\Bigr)$, which is the diagonal entry of $\bS^{(k)}(t_h, t_s)$. When $t_h = t_s$, $\bS^{(k)}(t_h, t_h)$ is the diagonol entry of $\bS^{(k)}$. Then the pooled within-class covariance matrix is $\mathbf{S}_p \in \Re^{pT \times pT}$, $\mathbf{S}_p = \frac{1}{\sum_{k = 1}^G (n_k-1)} \left(\sum_{k = 1}^G (n_k -1) \mathbf{S}^{(k)}\right)$. Given $\bSb$ and $\bSp$, LDA aims to find the solution to the optimization problem for the first discriminant matrix $\bbeta_1 \in \Re^{pT \times 1}$:

{\setstretch{1.0}\begin{equation}
\label{eqn:optimization}
\max_{\bbeta_1} \bbeta_1^{\smt} \bSb \bbeta_1 ~\mbox{subject to}~ \bbeta_{1}^{\smt} \bSp \bbeta_1 =1, ~\bbeta_1^{\smt} \bSp \bbeta_k = 0.
\end{equation}}

Then, we find the next discriminant matrix $\bbeta_k, \bbeta_k^{\smt} \bSp \bbeta_{l} = 0, k \ge 2, k \ne l$ by solving the LDA optimization problem after projecting data onto the orthogonal complement of $[\bbeta_1, \ldots, \bbeta_{k-1}]$. In other words, we deflate the data by obtaining $\mathbf{X}_{new}(t) = \mathbf{X}(t)P^{\perp}$, where $P^{\perp}$ is the projection matrix onto the orthogonal complement of $[\bbeta_1, \ldots, \bbeta_{k-1}]$, and we use the deflated data as the LDA objective for the subsequent discriminant function. If $y_i(t) = y_i$ so that class membership is time-independent, then we solve the optimization problem $\max_{\bbeta} \bbeta^{\smt} \bSb \bbeta ~\mbox{subject to}~ \bbeta^{\smt} \bSp \bbeta = 1$, for all time points simultaneously.

Without loss of generality, assume that $y_i$ is time-dependent. One can show, using Langragian multipliers, that the first discriminant function, $\bbeta_1$, can be obtained from solving the generalized eigenvalue-eigenvector system: $\bS_b \bbeta_1 = \lambda_1 \bS_p\bbeta_1$ for $(\widehat{\lambda}_1,\widehat{\bbeta}_1)$, which are the first eigenvalue-eigenvector pair of $\bSp^{-1}\bSb$ for $\bSp \succ 0$. Subsequent eigenvectors could be obtained such that $\bbeta_l^{\smt}\bS_p\bbeta_k = 0, l \ne k$. Let $\mathcal{M}=\bS_p^{{-1/2}}\bS_b\bS_p^{{-1/2}}$ and $(\widetilde{\lambda_k}, \widetilde{\gamma_k})$ be its eigenvalue-vector pair, which are solutions to the eigenvalue system: $\mathcal{M}\bgamma_k = \lambda_{k}\bgamma_k$, with $\gamma_l^{\smt}\gamma_k = 0, l \ne k$. It can be shown that the eigenvectors $\widetilde{\lambda}_k$ of $\mathcal{M}$ are the same as $\widehat{\lambda}_k$ (that is, $\widetilde{\lambda}_k = \lambda_k$) and the eigenvectors $\widetilde{\bgamma}_k = \bS^{{1/2}}_{p}\widehat{\bbeta}_k$. We choose to work with the eigenvalue-vector system instead of the generalized eigenvalue-vector system for computational ease. 

While our functional LDA framework accounts for inherent time dependency through the within-class and between-class covariance matrices, it also accommodates the assumption of time independence (i.e., no correlation between different time points). This simplification can be particularly advantageous for improving computational efficiency and in situations where temporal variation in discriminatory patterns is minimal or not of primary concern. Under this assumption, let $\bSb \in \Re^{pT \times pT}$ be a covariance function that collects $\bS_b(t_h)$ on the diagonals and $\mathbf{0}_{p \times p}$ on the off diagonals for all $T$ time points. Here, we assume that the between-class covariance matrices $\bSb(t_h)$ and $\bSb(t_h')$ are independent. Similarly, we define $\bSp \in \Re^{pT \times pT}$ be a covariance function that collects $\bS_p(t_h)$ on the diagonals and $\mathbf{0}_{p \times p}$ on the off diagonals for all $T$ time points. Here, we assume that the pooled within-class covariance matrices $\bS_p(t_h)$ and $\bS_p(t_h')$ are independent. Given $\bSb(t_h)$ and $\bSp(t_h)$, the LDA aims to find the solution to the optimization problem for the first discriminant function $\bbeta_1(t_h) \in \Re^{P \times 1}$ for each time point $t_h$ independently:

{\setstretch{1.0}\begin{equation}
\label{eqn:optimization}
\max_{\bbeta_1(t_h)} \bbeta_1^{\smt}(t_h) \bSb(t_h) \bbeta_1(t_h) ~\mbox{subject to}~ \bbeta_{1}(t_h)^{\smt} \bSp (t_h) \bbeta_1(t_h) =1, ~\bbeta_1^{\smt}(t_h) \bSp (t_h) \bbeta_k(t_h) = 0.
\end{equation}}

\subsection{Multivariate Sparse Functional LDA}

In the high-dimensional setting where $n \ll p$, the first discriminant function $\widetilde{\bgamma}_1$ is a weight vector of all the features available in $\bX(t)$. These coefficients are usually not zero (i.e., not sparse), making interpreting the discriminant function challenging. To enhance interpretation, we follow the ideas in \citep{SafoBiomSELP}, impose a convex penalty on $\bgamma_1$ and solve a problem similar to the Dantzig selector \citep{Dantzig:2007} for \( \widehat{\bgamma}_1 \):
\begin{equation}\label{eqn:sparseopt}
\min_{{\bgamma}_1} \sum_{i=1}^p|\gamma_i| \quad \text{s.t.} \quad \|\mathcal{M}\widetilde{\bgamma}_1 - \widetilde{\lambda}_1\bgamma_1 \|_\infty \leq \tau,
\end{equation}
where $\|\bx\|_{\infty} = \max_i|x_i|$, for a random vector $\bx$. Here, we constrain the eigenvalue system \( \mathcal{M}\bgamma_1 = \lambda_1\bgamma_1 \) to be within a threshold \( \tau \). However, simply constraining the eigensystem leads to trivial solutions. To address this concern, we replace \( \bgamma_1 \) on the left-hand side (LHS) of the eigensystem problem \( \mathcal{M}\bgamma_1 = \lambda_1\bgamma_1 \) with \( \widetilde{\bgamma}_1 \), the corresponding non-sparse solution. \( \widetilde{\lambda}_1 \) is the eigenvalue associated with \( \widetilde{\bgamma}_1 \). The parameter \( \tau \) is a tuning parameter that controls the level of sparsity, which will be selected adaptively using cross-validation to maximize multiple performance metrics including accuracy, balanced accuracy, F-1 score, precision, recall, and Matthew's Correlation Coefficient (MCC). The final features selected for the first discriminant function will be those with nonzero coefficients in at least 70\% of the time points. The dimension of $\bS_p$ makes the computation of $\bS_p^{{-1/2}}$ expensive. We proceed in two ways for computational efficiency. \textit{First}, for $p > n$, we solve the sparse discriminant function to obtain eigenvalues and eigen vectors using the ideas from \citep{HastieTrevor:2004}. Here, we avoid inverting the $pT \times pT$ matrix $\bS_p^{{-1/2}}$, instead,
we invert a $n \times n$ matrix. In this way, we reduce the $p$ variables to $n < p$ variables, then solve the $n$-dimensional optimization problem, and the solution can be transform back to $pT$ dimensions. This is also true for $n > p$. \textit{Second}, in the scenario where dataset has large $n$ and $p$, the computational storage of $\bS_p^{{-1/2}}$ will increase (an example of large simulation can be found in \textbf{Supplementary Material}). In this case, it is optional to assume no correlation between different time points dependents on computational resources availability. Thus, $\bS_p$ is block diagonal, and we can solve for $\bgamma_k(t_h)$ for each $t_h$, that is, we obtain $\widehat{\bgamma}_1(t_h)$, where we now invert $\bS_p(t_h)$. The performance of time-dependent and time-independent approach both shows competitive. This computation can be parallelized for computational speed. 

At convergence, we form the time-dependent discriminant scores \( \widehat{\bGamma}_1 \in \mathbb{R}^{p \times T} \), which collect the estimated \( \widehat{\bgamma}_1 \) for all time points. For classifying test data, we use the nearest centroid algorithm. Specifically, we project the training data tensor (\( \bX \in \mathbb{R}^{n \times p \times T} \)) and the testing data tensor (\( \bX_{\text{test}} \in \mathbb{R}^{n \times p \times T} \)) onto the discriminant matrix \( \widehat{\bGamma}_1 \), and assign a test sample to the class whose centroid is closest to the test sample. If the response \( y_i \) is time dependent, we classify at each time point; otherwise, we use all time points to determine the predicted class for the test sample. For subsequent discriminant functions, we deflate the data and proceed to solve equation (\ref{eqn:sparseopt}) for subsequent sparse discriminant functions. 

The optimization problem relies on the tuning parameter \(\tau\), which must be carefully selected to avoid trivial solutions. To balance sparsity and performance, a range of \(\tau\) values is selected, with eight grid points between predefined upper and lower bounds, ensuring that the sparsity rate falls within a target range. After determining an appropriate \(\tau\) range, cross-validation is used to select the optimal \(\tau\) by maximizing a combined performance metric, which includes accuracy, balanced accuracy, F-1 score, precision, recall, and Matthew's Correlation Coefficient (MCC). Algorithms of $\tau$ range selection, tuning details, and cross-validation can be found in \textbf{Supplementary Material}.

Our method provides unique advantages that distinguish it from current approaches, significantly improving biomedical research. \textit{First}, it is applicable to multivariate longitudinal data. This allows us to investigate how a specific feature changes over time, while considering the influence of other features. \textit{Second}, the FDA technique is highly flexible and can handle uneven time points effectively, as it can interpolate observations at any time point $t = 1,\ldots,T$. \textit{Third}, it can identify profiles that discriminate between groups over time, enhancing interpretation. \textit{Fourth}, through the discriminant scores, one can clearly pinpoint the exact time points where groups exhibit the most significant differences visually (e.g., Figure 2). Furthermore, a formal test, such as the Kolmogorov-Smirnov test, can be employed to statistically determine whether the distributions of the two groups differ based on the discriminant scores.

\section{Simulations}
\label{sec:simulation}

We conduct extensive simulations to evaluate the effectiveness of the proposed method and compare it to several existing approaches. Our simulation design follows the ideas of \cite{gardner2021linear} and \citep{cao2012simultaneous} with extensions to multi-class and high-dimensional settings. We consider both binary ($K=2$) and multi-class ($K=3$) classification problems, with varying sample sizes and number of variables, to capture scenarios when $n \ll p$ and $n \gg p$. We also generate simulations with or without subject-specific temporal effect, to capture time-dependent and time-independent scenarios. In all scenarios, we assume that few features are signals and differ between classes, while other features have little or no differences between the groups. An illustration of selected and non-selected features can be found in \textbf{Supplementary Material}. We simulate data with $T = 40$ time points, different mean separation functions, and varying group differences. We simulate 100 Monte Carlo replicates and report average performance metrics.

We compare our proposed method with the following methods: LDA for multiple functional data [FLDA] \citep{gardner2021linear}, penalized LDA [PLDA] \citep{witten2011penalized}, and functional principal component analysis \citep{yao2005functional} followed by penalized LDA [FPCA + PLDA]. FLDA is a functional LDA method for multivariate data that is only applicable to binary classification problems and is not capable of selecting features. PLDA is a multivariate LDA method for cross-sectional data that is capable of selecting features. Since PLDA is only applicable to cross-sectional data, we implement it at each time point, obtain the predicted class for that time point, and use majority voting for the overall predicted class. For FPCA, we obtain the first principal component for each feature and then implement PLDA on the combined functional principal component scores for all features. Implementing PLDA after FPCA will allow us to choose the functional principal component (FPC) score of a feature that can discriminate between groups. Thus, for FPCA + PLDA, we consider a feature selected and able to discriminate between the groups over time if their FPC scores are selected by PLDA. 

In each simulation example, we generate 100 independent training and testing sets. We compare the effectiveness of the methods in terms of feature selection (ability to select signal features that discriminate the groups over time while ignoring noise features) and prediction performance, using sensitivity, specificity, and F-1 score. 

Due to space limitations, we present the comparison results for the multi-class classification problem in the main article, and refer to the \textbf{Supplementary Material} for details on the binary classification problem and high-dimensional settings.

\subsection{Multi-Class Problem}
\begin{figure}[h]
    \centering
        \includegraphics[width=\textwidth]{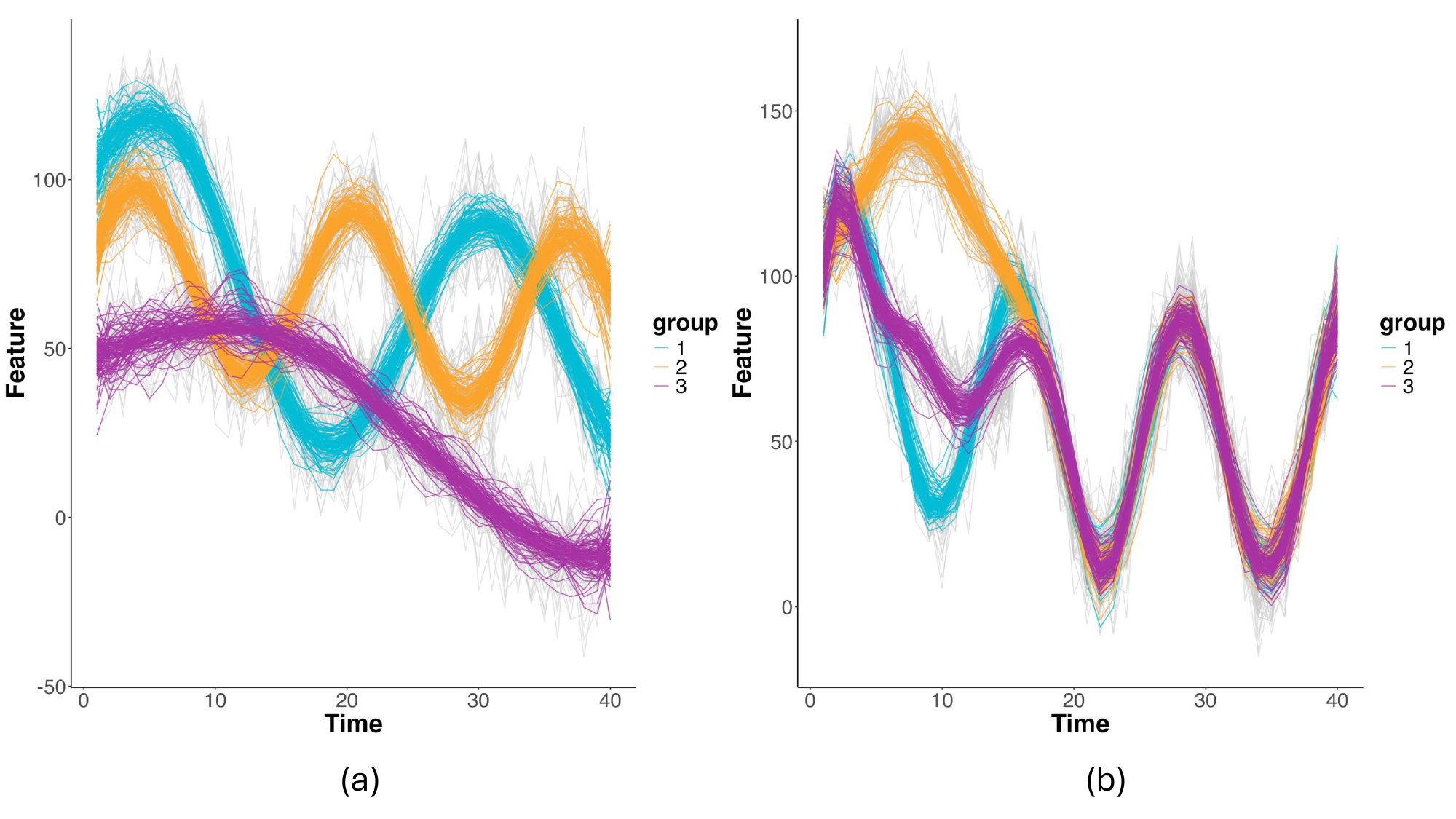}
        \caption{Illustration of Multi-Class Simulation Datasets. Gray line represents the raw simulation and colored lines are B-spline smoothed estimations for each group: (a) groups separated at all time points; (b) groups only separated between partial time range.}
        \label{fig:simulation-3-allthree}
\end{figure}

We generate a set of curves over the time interval $T = [1,40]$ for $n_1$ samples belonging to group 1, $n_2$ samples in group 2, and $n_3$ samples in group 3. We generate $p$ features of function curves for each sample. The first 10\% of the features can discriminate between the three classes (Figure \ref{fig:simulation-3-allthree}). Following \citep{gardner2021linear} and \citep{cao2012simultaneous}, each set of functions was generated by: 
\[
x_{ij}(t) = \lambda \times \delta + \eta_{0j} + \eta_{1j}t + \eta_{2j}t^2 + \eta_{3j}t^3 + \eta_{4j}t^4 + \eta_{5j}\sin(\eta_{6j}t) + \rho STE_{ij}(t) + \epsilon_{ij},
\]
where $\eta_{0j}, \eta_{1j}, \eta_{2j}, \eta_{3j}, \eta_{4j}, j = 1,\ldots,p$, are generated from the least squared fit of a fourth-degree polynomial to $(x_{1j}^{*}, y_{1j}^{*}), (x_{2j}^{*}, y_{2j}^{*}), (x_{3j}^{*}, y_{3j}^{*}), (x_{4j}^{*}, y_{4j}^{*}), (x_{5j}^{*}, y_{5j}^{*}), (x_{6j}^{*}, y_{6j}^{*})$. Here $x_1^* = 0$, $x_6^* = 10, x_k^* \sim \text{unif}(0,10), k = 2, 3, 4, 5$ and $y_k^* \sim \text{unif}(50,100), k = 1,\ldots,6$. $\eta_{5j} = \max(\eta_{0j} + \eta_{1j}t + \eta_{2j}t^2 + \eta_{3j}t^3 + \eta_{4j}t^4) - \min (\eta_{0j} + \eta_{1j}t + \eta_{2j}t^2 + \eta_{3j}t^3 + \eta_{4j}t^4)$ under the interval $t = 1, 2, \ldots, 40$, and $\eta_{6j} \sim U(0,10)$, $\epsilon_{p,ij} \sim N(0, \sigma^2)$. The value of $\delta$ represents the separation between the two groups, $\delta = 0$ for group 1, $\delta = 500$ for group 2, and $\delta = 1000$ for group 3. The indicator $\lambda$ represents which group has a higher mean function, $\lambda = 1$ for group 1, and $\lambda$ is randomly chosen from 1 or -1 for the other two groups. We also consider subject-specific temporal effect $\rho STE_{ij}(t) = \rho \sum_{k=1}^2 \xi_{ijk}\psi_k(t)$, where $\rho = 0, 1$, $\xi_{ijk} \sim N(0, 1)$, $k = 1, 2$, $\psi_1(t) = -2\cos\{\pi(t-1/2)\}$, and $\psi_2(t) = \sin\{\pi(t-1/2)\}$. We assume that such a subject-specific temporal effect is independent across each feature. We consider simulations for both with subject-specific temporal effect (i.e., $\rho = 1$) and without subject-specific temporal effect (i.e., $\rho = 0$) to assess the model robustness under varying assumptions about subject-specific temporal effects.

We consider five different scenarios: (1) the three groups are distinct across all time points, exhibiting similar average trends; (2) the three groups diverge only within a specific time window, from the time point [5, 15], displaying different patterns; (3) the three groups show separation over a fixed interval of 10 time points, though the exact time frame varies; (4) the three groups are distinct during a random occurring period, with the duration of separation varying between 5 and 40 time points. In this case, the groups differ in both their mean functions and temporal patterns. For cases (1) through (4), we examine scenarios without subject-specific temporal effects (i.e., $\rho = 0$). Finally, (5) considers the same setting as case (2) -- group divergence within time points 5 to 15 -- but includes additional subject-specific temporal effects (i.e., $\rho = 1$). 
Figure \ref{fig:simulation-3-allthree} illustrates some cases in which the gray lines represent the observed data, while the colored lines reflect the smoothed B-spline estimations. The groups are differentiated by colors. In Figure \ref{fig:simulation-3-allthree} (a), the three groups consistently display distinct mean values across all time points, although their patterns remain similar. However, group differences do not always persist at all time points. In more realistic scenarios, groups may differ only during specific time intervals. In the second case, the groups are largely identical in their mean functions, but their patterns diverge during the period [5, 15], as shown in Figure \ref{fig:simulation-3-allthree} (b). In the third case, the three groups differ over a fixed time span of 10 points, but the location of this time window varies. In the fourth scenario, the groups separate over a randomly occurring period, and the duration of this separation also varies. The group differences may manifest through either dissimilar mean functions or similar means with divergent patterns. 

In this section, we assume that the features can differentiate between all three groups. However, in other scenarios, the features that separate group 1 from group 2 may differ from those that distinguish group 2 from group 3. A detailed simulation study and model comparison for these scenarios are provided in the \textbf{Supplementary Material}.

\subsection{Multi-Class Simulation Results and Comparison}

\begin{table}[h]
\caption{Multi-Class Classification (left) and feature selection (1st discriminant vector) test metrics for simulated data based on 100 repetitions. The highest metrics in each case are in bold.}
    \label{tab:simresults-multi}
    \centering
    \resizebox{\columnwidth}{!}{\begin{tabular}{c|c|ccc | ccc}
    \hline
    \hline
      &       &\multicolumn{3}{c|}
     {Classification} & \multicolumn{3}{c}{Feature Selection} \\ \cline{3-8}
    Case &  Model & W.Sens & W.Spec & W.F-1 & Sens & Spec & F-1\\
    \hline
    Different in   &   MFLDA-D   &  \textbf{1.00} &  \textbf{1.00} &  \textbf{1.00} & \textbf{1.00} & \textbf{1.00} &  \textbf{1.00}\\
    all [0,40]   &   MFLDA-I   &  \textbf{1.00} &  \textbf{1.00} &  \textbf{1.00} & \textbf{1.00} & \textbf{1.00} &  \textbf{1.00}\\
       &   PLDA   &  0.99 & 0.99  &  0.99 &  0.42 &  0.70 &  0.25 \\
       &   FPCA+PLDA   & 0.96  & 0.98 &  0.96 &  0.00 &  \textbf{1.00} &  -\\
    \hline
    Different only   &   MFLDA-D   &  \textbf{1.00}  &  \textbf{1.00} &  \textbf{1.00} &  \textbf{1.00}  & \textbf{1.00}  &  \textbf{1.00}\\
    in [5,15]   &   MFLDA-I   &  \textbf{1.00}  &  \textbf{1.00} &  \textbf{1.00} &  0.90  & 0.93  &  0.72\\
       &   PLDA   &  0.82  &  0.91 &  0.82 &  0.66  &  0.43 &  0.25\\
       &   FPCA+PLDA   &  0.79 & 0.89  & 0.79 &  \textbf{1.00} &  0.00 &  -\\
    \hline
    Different in   &   MFLDA-D   & \textbf{1.00}  & \textbf{1.00}  & \textbf{1.00} & \textbf{1.00}  &  0.90 &  \textbf{0.99}  \\
    random 10 time   &   MFLDA-I   & \textbf{1.00}  & \textbf{1.00}  & \textbf{1.00} & 0.90  &  0.97 &  0.82  \\
     points  &   PLDA   &  0.98 & 0.99  &  0.98 &  0.27 &  0.83 &  0.20 \\
       &   FPCA+PLDA   & 0.74  & 0.87  &  0.74 &  0.00 &  \textbf{1.00} &  -\\
    \hline
    Different in random   &   MFLDA-D  & \textbf{1.00}  & \textbf{1.00}  &  \textbf{1.00} & 0.99  & \textbf{1.00}  &  \textbf{0.99} \\
    time and random   &   MFLDA-I   & \textbf{1.00}  & \textbf{1.00}  &  \textbf{1.00} & \textbf{1.00}  & 0.82  &  0.56 \\
     length of time   &   PLDA   & 0.99  &  0.99 & 0.99 & 0.31  & 0.90 & 0.35 \\
      &   FPCA+PLDA   & 0.94  & 0.97  &  0.94 &  0.00 &  \textbf{1.00} &  -\\
      
    \hline
    Different only   &   MFLDA-D   &  \textbf{1.00} & \textbf{1.00} & \textbf{1.00} & \textbf{0.77} & 0.90 & \textbf{0.86}\\
    in [5,15]   &   MFLDA-I   & 0.98 & 0.99 & 0.97 & 0.12 & \textbf{1.00} & 0.22 \\
    with $\rho = 1$   &   PLDA   & 0.89 & 0.94 & 0.88 & 0.26 & 0.81 & 0.18\\
       &   FPCA+PLDA   & \textbf{1.00} & \textbf{1.00} & \textbf{1.00} & 0.10 & \textbf{1.00} & 0.18 \\
    
    \hline
    \hline
    \end{tabular}}

\end{table}   

Since the features that differentiate the groups are identical, both MFLDA and other competing methods are able to identify the first discriminant vector but struggle to extract the second. Table \ref{tab:simresults-multi} summarizes the classification and feature selection results. We denote MFLDA-D as the method that accounts for time-dependent between- and within-class covariances, and MFLDA-I as the one assuming time-independence in these structures. Across all scenarios, MFLDA-D consistently outperforms or matches other methods, achieving perfect classification (weighted sensitivity, weighted specificity, weighted F-1 score = 1.00) in nearly every settings. Notably, it maintains excellent performance even in challenging cases involving randomly occurring time windows and varying separation lengths, underscoring the value of explicitly modeling time-dependent covariance structures in dynamic data. MFLDA-I also performs strongly in classification, with scores identical to MFLDA-D in many scenarios. This indicates that MFLDA is highly effective at accurately distinguishing between multiple classes, making it a robust and reliable method for complex classification tasks. However, its performance slightly declines in the [5,15] with $\rho = 1$ setting, where time-invariant assumptions are violated—highlighting the advantage of MFLDA-D under temporal heterogeneity.

In comparison, PLDA performs well but falls slightly short of MFLDA's consistency. With weighted sensitivity, weighted specificity, and weighted F-1 scores ranging from 0.82 to 0.99, PLDA demonstrates strong classification capabilities, although it may be slightly less reliable in more complex or variable scenarios. Although still a viable option, its slight variability in performance suggests that it may not always be the best choice when high accuracy and consistency are critical. The performance of FPCA+PLDA is suboptimal. Its weighted sensitivity, weighted specificity, and weighted F-1 scores range from 0.74 to 1.00, indicating struggles with maintaining high accuracy, particularly when faced with varying conditions such as random time points or differing time lengths. This suggests that FPCA+PLDA may be less effective when a high level of generalization and accuracy is required, especially in more complex datasets.

In terms of feature selection, MFLDA-D remains the top performer, particularly in more complex and time-varying scenarios. It achieves high F-1 scores (often $\geq 0.99$) and balanced sensitivity and specificity, even under irregular time structures and subject-specific temporal effects ($\rho STE(t)$). This highlights the effectiveness of MFLDA-D in identifying and selecting the most relevant features that contribute to the discriminant, which is crucial for producing accurate and interpretable classification results. MFLDA-I also outperforms competing methods. It achieves higher scores on the sensitivity, specificity, and F-1 metrics. However, MFLDA-I shows noticeable degradation in the $\rho = 1$ case, where its feature selection F-1 drops to 0.22, with very low sensitivity (0.12). This again reflects its vulnerability when temporal dynamics are present but unmodeled. PLDA shows moderate performance in feature selection, with sensitivity and specificity scores that fluctuate depending on the scenario. Sensitivity scores range from 0.27 to 0.66, specificity scores from 0.43 to 0.90, and F-1 scores from 0.20 to 0.35, reflecting some inconsistency. Although PLDA can sometimes effectively select relevant features, it may not do so consistently, particularly in more challenging scenarios. This makes it less reliable for tasks where precise selection of features is essential. FPCA+PLDA struggles the most, often failing to identify relevant features effectively. Sensitivity scores are as low as 0.00 in several cases, and while specificity scores are consistently high, this does not compensate for its poor overall performance in feature selection. This suggests that FPCA+PLDA might not be the best choice in scenarios where precise and reliable feature selection is essential for classification success. 

\begin{figure}[h!]
    \centering
        \includegraphics[width=\textwidth]{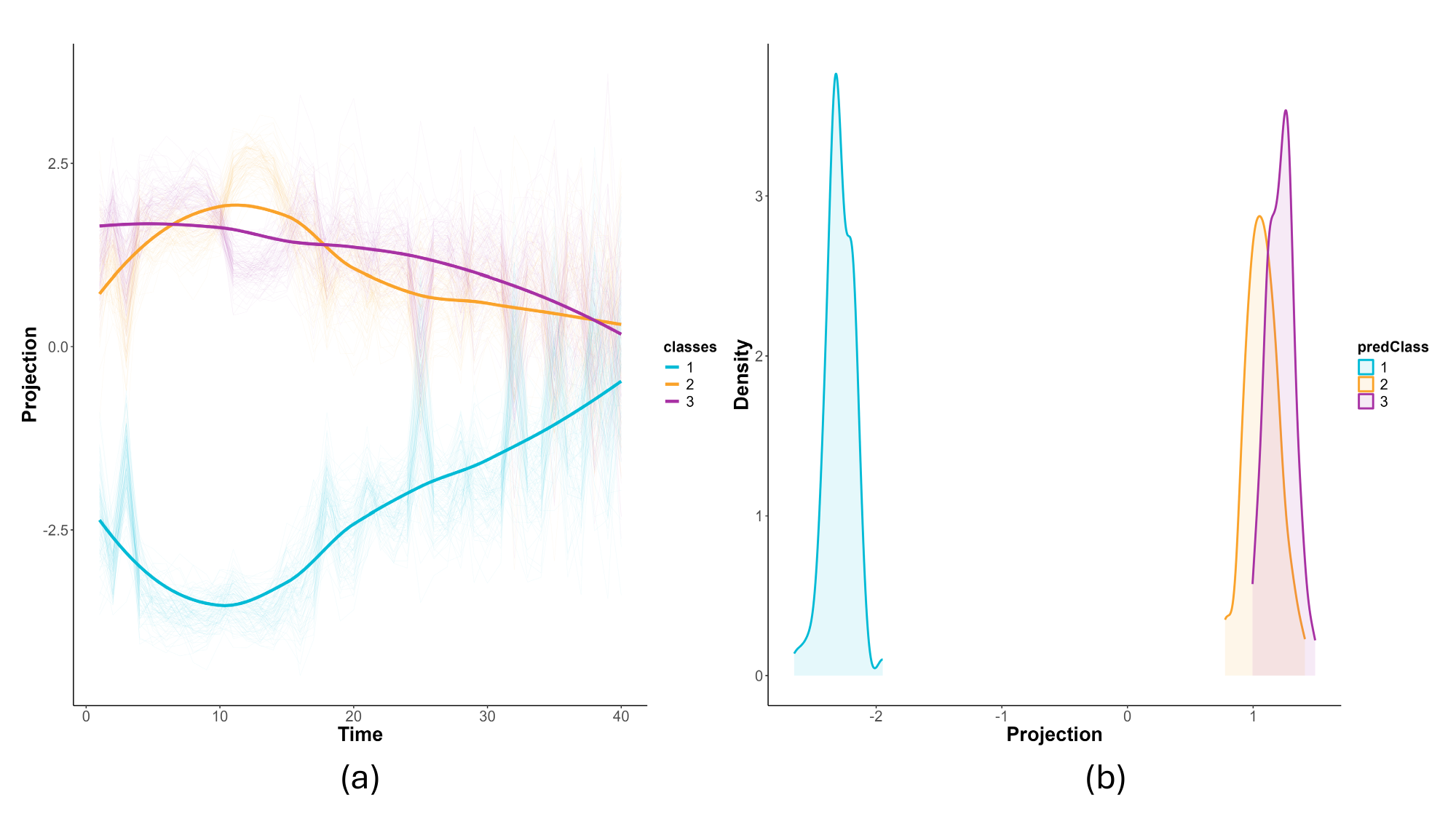}
        \caption{Discriminant Plot of Multi-Class: (a) Discriminant plot for first discriminant vector. Thin lines represent projection for each individual and thick lines are LOESS curves colored by classes; (b) Density plot of first discriminant vector.}
        \label{fig:simulation-multi-result}
\end{figure}

Figure \ref{fig:simulation-multi-result} presents a discriminant plot for better visualization. In panel (a), the plot includes a smoothed LOESS curve, highlighting the separation of the three groups between the time interval [5, 15]. Beyond this range, the groups become less distinguishable. The plot aligns with the results of Scenario 5. Furthermore, Figure \ref{fig:simulation-multi-result} (b) presents a projection density plot, where we observe that three classes are quite distinguishable. 

\section{Real Data Application}
\label{sec:IBD}

\subsection{Dataset and Preprocessing}

The iHMP IBD dataset includes data from 130 individuals, collected over a 50-week period. We choose cubic spline with 4 interior knots. We then exclude 30 individuals who do not have a minimum number of 8 time points. Since MFLDA is not speicifc to microbiome data, we preprocessed the IBD data (e.g., exclude pathways with low variations, and apply central log ratio transformation) to overcome the zero inflation and compositional nature of microbiome data \citep{lloyd2019multi}. The number of microbial pathways remaining for the metagenomics is 1,622. We apply a Linear Mixed Model (LMM) \citep{liu2024incorporating} to select the top 200 significant metagenomics pathways for further analysis. A detailed description of data preprocessing can be found in \textbf{Supplementary Material}. Due to the small sample size of the IBD dataset, we use 5-fold cross-validated metrics to minimize data variability. Microbiome pathways selected by MFLDA are all pathways selected by all 5 folds.

\subsection{Multi-Class Outcome Results (CD vs. UC vs. Non-IBD)}

We apply our proposed and competing methods to identify microbial pathways that discriminate between IBD status (non-IBD, UC, and CD). We consider binary discrimination (IBD vs Non-IBD) in the \textbf{Supplementary Material}. There are more subjects in the CD group (48 individuals) than in the UC group (26 individuals) and the non-IBD group (26 individuals).


\begin{table}[h]
\caption{Classification Metrics for Predicting 3 IBD Groups based on 5-Fold Cross-Validation. The highest metrics are in bold.}
    \label{tab:IBD3results}
    \centering
    \begin{tabular}{r|ccccc}
    \hline
    \hline
    Metrics & MFLDA-D & MFLDA-I & PLDA & FLDA & FPCA+PLDA\\
    \hline
    Weighted Sensitivity & \textbf{0.540} &  0.530 & 0.445 & - & 0.379\\
    Weighted Specificity & 0.798 &  \textbf{0.801}  & 0.585 & - &  0.615 \\
    Weighted F-1 Score & \textbf{0.523} &  0.520  & 0.415 & - & 0.380 \\
    Accuracy  & \textbf{0.540} &  0.530  & 0.445 & - &  0.379\\
    Balanced Accuracy   & \textbf{0.669} & 0.663  & 0.515 & - & 0.497\\
    MCC & \textbf{0.357} &  0.342  & 0.047 & - & -0.008\\
    \hline
    \hline
    \end{tabular}
\end{table}

Table \ref{tab:IBD3results} compares the classification performance of MFLDA to other methods in weighted sensitivity, weighted specificity, weighted F-1 score, accuracy, balanced accuracy, and MCC. Both MFLDA-D and MFLDA-I deliver comparable classification performance. Compared to other methods, they consistently stand out with higher weighted sensitivity, weighted specificity, and MCC. With both higher sensitivity and specificity, our method is better at identifying true positives while minimizing false positives. A high F-1 score and MCC further indicate their superior overall performance. 

MFLDA-D selects 76 and MFLDA-I selects 59 microbial pathways that distinguish between individuals with UC, CD, and those without IBD [non-IBD], demonstrating strong feature selection performance. In contrast, PLDA, which works independently at each of the 50 time points, chooses all pathways at 17 time points. The two-step FPCA+PLDA approach also selects all pathways. However, despite that many microbial pathways are selected, neither PLDA nor FPCA+PLDA showed any improvement in classification performance compared to MFLDA. Meanwhile, FLDA cannot be applied to three-class classification problems.

The performance of multi-class classification is generally lower than that of binary classification, indicating the challenges in distinguishing between the multiple IBD groups over time. However, our proposed method demonstrates better performance compared to other approaches, underscoring its potential for effective multivariate functional discrimination.

\subsubsection{Microbiome Pathways Selected by SFLDA}

We applied both MFLDA-D and MFLDA-I to the IBD dataset. While the microbiome pathways selected by the two methods show some overlap, they also differ. Due to space constraints, we focus on discussing the pathways identified by MFLDA-D and those shared by both methods. MFLDA-D selects 76 microbiome pathways that can distinguish between subjects with UC, CD, and non-IBD. Figure \ref{fig:pathways_vis_3_dep} is an animation of all selected microbial pathways, with colored lines showing the average feature abundance by group. In some pathway profiles, CD and UC exhibit similar patterns [e.g., Pathway\_45, Pathway\_50] and are separated from non-IBD, which may have different means or opposite patterns. In some cases, non-IBD is distinguishable from UC but not CD [e.g., Pathway\_51, Pathway\_191]. MFLDA-D effectively identifies pathways that distinguish these groups.

\begin{figure}
\centering
\animategraphics[loop, controls, width=15cm]{2}{animation_ibd3_dep/pathway}{1}{76}
\caption{Selected Features for CD vs. UC vs. Non-IBD Status. Lines represented average time series curves of standardized feature abundance for each IBD status group. Highlighted pathways are mentioned in main text. Click to start the figure animation and use controls to navigate, start, and stop.}
\label{fig:pathways_vis_3_dep}
\end{figure}

Certain microbiome processes related to the human gut microbiota, such as carbohydrate metabolism, Coenzyme A biosynthesis and amino acid metabolism, can help distinguish IBD subtypes (UC and CD). Using our method, we identified relevant microbiome pathways. The carbohydrate metabolism process, particularly associated with the \textit{Alistipes putredinis}, has anti-inflammatory effects, highlighting the distinct metabolic environments in UC \citep{nomura2021bacteroidetes, wu2024changes}. \textit{Ruminococcus bicirculans}, plays a significant role in both Coenzyme A biosynthesis and amino acid metabolism, depletes especially in CD compared with non-IBD \citep{mayorga2022intercontinental}. Discussion of 59 microbiome pathways selected by MFLDA-I can be found in the \textbf{Supplementary Material}. Among the 41 common pathways selected by both MFLDA-D and MFLDA-I (Table \ref{tab:var-compare-mflda}), tryptophan metabolism and mucin degradation are particularly relevant to IBD pathogenesis. Tryptophan metabolism influences inflammation and is associated with activity of IBD \citep{nikolaus2017increased}. Concurrently, increased abundance of \textit{Akkermansia muciniphila}, a mucin-degrading bacterium, reflects compromised mucus layer integrity, facilitates immune activation \citep{zheng2023role}. These pathways help differentiate microbial activity and inflammation in UC and CD from each other and from non-IBD conditions. Other selected pathways could be further studied for their potential roles in IBD.

The results of both MFLDA-D and MFLDA-I feature selection also emphasize the crucial role of vitamin intake in IBD (vitamin related pathways are highlighted in Table \ref{tab:var-compare-mflda}). They shed light on how different vitamins, particularly those in the B-complex group—like thiamin (B1) \citep{fernandez1989vitamin}, riboflavin (B2) \citep{levit2018effect}, pantothenate (B5) \citep{kuroki1993multiple} and folate (B9) \citep{ratajczak2021does}—affect IBD status. These vitamins are important for cellular metabolism, immune function, and the management of inflammation. Although most studies have focused on individual vitamins in IBD, our findings emphasize the importance of considering multiple vitamins together, as they can collectively affect IBD risk. These findings reveal a complex relationship between microbial pathways and vitamin metabolism and suggest  that a more integrated nutritional approach may be important to effectively manage IBD.

\begin{table}[h]
\caption{Microbiome Pathways (shortened names due to space constraint) Selected by MFLDA-D and MFLDA-I: common pathways selected are in red (41 pathways), vitamin related pathways are highlighted.}
    \label{tab:var-compare-mflda}
    \centering
    \begin{tabular}{p{7.4cm}|p{7.4cm}}
    MFLDA-D (76 pathways selected) & MFLDA-I (59 pathways selected)\\
    \hline
    \hline
    \fontsize{7.5}{7.5}\selectfont\label{var-multi-MFLDA-I}

\textcolor{red}{\textbf{UNINTEGRATED - Akkermansia - Muciniphila}},
Ruminococcus - torques, 
Roseburia - hominis, 
Roseburia - inulinivorans, 
Roseburia - inulinivorans - CAG - 15, 
Ruminococcus - bicirculans, 
\highlight{yellow}{red}{\textbf{N10 - Formyl - Tetrahydrofolate - Unclassified}}, 
\textcolor{red}{\textbf{Glycolysis - III - Glucose}}, 
chorismate - biosyn - I, 
branched - aa - biosyn - Ruminococcus - bicirculans, 
branched - aa - biosyn, 
\hl{coA - biosyn - II - Ruminococcus - bicirculans}, 
\highlight{yellow}{red}{\textbf{CoA - Biosynthesis - Unclassified}}, 
\hl{coA - biosyn - I - Ruminococcus - bicirculans}, 
aromatic - aa - biosyn, 
dTDP - rhamnose - biosyn - I - Alistipes - putredinis - CAG - 67,
gluconeogenesis - I, 
ornithine - biosyn, 
glycolysis - I, 
\textcolor{red}{\textbf{Histidine - Biosynthesis}}, 
isoleucine - biosyn - I - Ruminococcus - bicirculans, 
isoleucine - biosyn - I, 
\textcolor{red}{\textbf{Formaldehyde - Assimilation}}, 
\textcolor{red}{\textbf{Incomplete - Reductive - TCA}},
\highlight{yellow}{red}{\textbf{Phosphopantothenate}}, 
\highlight{yellow}{red}{\textbf{Pantothenate - CoA}}, 
peptidoglycan - biosyn - I, 
\textcolor{red}{\textbf{Glycolysis - IV - Plant}}, 
\hl{coA - biosyn - III - Ruminococcus - torques}, 
\textcolor{red}{\textbf{Histidine - Degradation - Alistipes}}, 
isoleucine - biosyn - III - Ruminococcus - bicirculans, isoleucine - biosyn - III, 
\textcolor{red}{\textbf{Unsaturated - Fatty - Acid - Oxidation}},
\textcolor{red}{\textbf{Sulfur - Oxidation - Acidianus}}, 
glycolysis - II, 
\textcolor{red}{\textbf{Glutamate - Glutamine}},
\textcolor{red}{\textbf{GDP - Mannose}}, 
AIR - biosyn - I, 
chorismate - biosyn - 3DHQ, 
\textcolor{red}{\textbf{Sucrose - Degradation}}, 
\highlight{yellow}{red}{\textbf{Menaquinol - 8}}, 
putrescine - biosyn - IV - Ruminococcus - bicirculans, 
UDP - MurNAc - pentapeptide - biosyn - II, 
UDP - MurNAc - pentapeptide - biosyn - I, 
\textcolor{red}{\textbf{Peptidoglycan - Enterococcus}}, 
stachyose - deg, 
\hl{queuosine - biosyn - Ruminococcus - torques}, 
\textcolor{red}{\textbf{PreQ0 - Biosynthesis}}, 
\highlight{yellow}{red}{\textbf{Thiazole - Unclassified}}, 
\highlight{yellow}{red}{\textbf{Thiamin - Salvage}}, 
\textcolor{red}{\textbf{TCA - Oxoglutarate}}, 
\textcolor{red}{\textbf{Pyruvate - Fermentation - Isobutanol}},
adenosine - rn - biosyn - Ruminococcus - bicirculans, 
guanosine - rn - biosyn - Ruminococcus - bicirculans, 
guanosine - rn - biosyn, 
\textcolor{red}{\textbf{Inosine - Biosynthesis - III}},
\textcolor{red}{\textbf{Inosine - Biosynthesis}}, 
\textcolor{red}{\textbf{Anaerobic - Metabolism - Invertebrates}},
\textcolor{red}{\textbf{Anaerobic - Metabolism}}, 
\textcolor{red}{\textbf{Mannan - Degradation}}, 
\textcolor{red}{\textbf{Dihydropterin - Chlamydia - CAG67}}, 
\textcolor{red}{\textbf{Gondoate - Anaerobic}}, 
purine - rn - deg - Ruminococcus - torques, 
\textcolor{red}{\textbf{Purine - Degradation - Eubacterium}},
\textcolor{red}{\textbf{Peptidoglycan - Maturation - Eubacterium}},
\textcolor{red}{\textbf{Gluconeogenesis - III}}, 
gluconeogenesis - III - unclassified, 
\hl{NAD - biosyn - I - Alistipes - finegoldii}, 
\highlight{yellow}{red}{\textbf{Flavin - Biosynthesis - Bacteria}},
\textcolor{red}{\textbf{Serine - Glycine - Biosynthesis - Alistipes}}, 
\highlight{yellow}{red}{\textbf{Thiamin - Diphosphate - Bacteroides}}, 
\highlight{yellow}{red}{\textbf{Thiamin - Diphosphate}}, 
\highlight{yellow}{red}{\textbf{Thiamin - Diphosphate - Eukaryotes - Blautia}}, 
\textcolor{red}{\textbf{Tryptophan}}, 
\textcolor{red}{\textbf{UDP - Acetyl - Glucosamine - Ruminococcus}}, 
\textcolor{red}{\textbf{Valine - Ruminococcus}}
   &   
    \fontsize{7.5}{7.5}\selectfont\label{var-multi-MFLDA-I}
\textcolor{red}{\textbf{UNINTEGRATED- Akkermansia- Muciniphila}}, 
UNINTEGRATED- Alistipes- Putredinis, 
UNINTEGRATED- Alistipes- Putredinis- CAG67, 
\hl{N10- Formyl- Tetrahydrofolate- Alistipes- CAG67}, 
\highlight{yellow}{red}{\textbf{N10- Formyl- Tetrahydrofolate- Unclassified}}, 
\highlight{yellow}{red}{\textbf{Glycolysis- III- Glucose}}, 
\hl{CoA- Biosynthesis- Ruminococcus}, 
\highlight{yellow}{red}{\textbf{CoA- Biosynthesis- Unclassified}},
dTDP- Rhamnose- Alistipes- CAG67, 
\textcolor{red}{\textbf{Histidine- Biosynthesis}}, 
\textcolor{red}{\textbf{Formaldehyde- Assimilation}}, 
\textcolor{red}{\textbf{Incomplete- Reductive- TCA}},
\highlight{yellow}{red}{\textbf{Phosphopantothenate}}, 
\highlight{yellow}{red}{\textbf{Pantothenate- CoA}}, 
\textcolor{red}{\textbf{Glycolysis- IV- Plant}}, 
CMP- 3- Deoxy- D- Manno, 
\textcolor{red}{\textbf{Histidine- Degradation- Alistipes}}, 
Pyruvate- Fermentation- Acetate- Blautia, 
Isoleucine- Biosynthesis- Alistipes, 
Isoleucine- Biosynthesis- Blautia, 
\textcolor{red}{\textbf{Unsaturated- Fatty- Acid- Oxidation}}, 
\textcolor{red}{\textbf{Sulfur- Oxidation- Acidianus}}, 
Sulfur- Oxidation- Blautia, 
\textcolor{red}{\textbf{Glutamate- Glutamine}}, 
\textcolor{red}{\textbf{GDP- Mannose}}, 
Cis- Vaccenate, Dihydropterin- Alistipes- CAG67, 
\textcolor{red}{\textbf{Sucrose- Degradation}}, 
\highlight{yellow}{red}{\textbf{Menaquinol- 8}}, 
Peptidoglycan- Mycobacteria- Alistipes, 
\textcolor{red}{\textbf{Peptidoglycan- Enterococcus}}, 
\textcolor{red}{\textbf{PreQ0- Biosynthesis}}, 
\hl{Thiazole- Bacteroides}, 
\highlight{yellow}{red}{\textbf{Thiazole- Unclassified}},
\highlight{yellow}{red}{\textbf{Thiamin- Salvage}}, 
\textcolor{red}{\textbf{TCA- Oxoglutarate}}, 
\textcolor{red}{\textbf{Pyruvate- Fermentation- Isobutanol}}, 
\textcolor{red}{\textbf{Inosine- Biosynthesis- III}}, 
\textcolor{red}{\textbf{Inosine- Biosynthesis}}, 
\textcolor{red}{\textbf{Anaerobic- Metabolism- Invertebrates}}, 
\textcolor{red}{\textbf{Anaerobic- Metabolism, Mannan- Degradation}},
\highlight{yellow}{red}{\textbf{Dihydropterin- Chlamydia- CAG67}},
\textcolor{red}{\textbf{Gondoate- Anaerobic}}, 
ADP- Glycero- Manno, 
\textcolor{red}{\textbf{Purine- Degradation- Eubacterium}},
\textcolor{red}{\textbf{Peptidoglycan- Maturation- Eubacterium}}, 
\textcolor{red}{\textbf{Gluconeogenesis- III}}, 
Glycolysis- VI- Metazoan, 
\highlight{yellow}{red}{\textbf{Flavin- Biosynthesis- Bacteria}}, 
\textcolor{red}{\textbf{Serine- Glycine- Biosynthesis- Alistipes}},
\highlight{yellow}{red}{\textbf{Thiamin- Diphosphate- Bacteroides}}, 
\highlight{yellow}{red}{\textbf{Thiamin- Diphosphate}}, 
\highlight{yellow}{red}{\textbf{Thiamin- Diphosphate- Eukaryotes- Bacteroides}}, 
\highlight{yellow}{red}{\textbf{Thiamin- Diphosphate- Eukaryotes- Blautia}}, 
\textcolor{red}{\textbf{Tryptophan}}, 
\textcolor{red}{\textbf{UDP- Acetyl- Glucosamine- Ruminococcus}}, 
\textcolor{red}{\textbf{Valine- Ruminococcus}}, 
Valine- Unclassified\\
    \end{tabular}
\end{table}

\section{Conclusion}
\label{sec:conclusion}

Multivariate longitudinal data are important in biomedical research and offer deep insight into the dynamics of complex disease over time. However, there is a gap in the literature in terms of methods that can effectively analyze and interpret such data, particularly for classification, while also selecting variables that differentiate between classes over time. To address these limitations, we introduced MFLDA ( Multivariate Functional Linear Discriminant Analysis), a novel method capable of discriminating between two or more classes while providing interpretable feature selection assuming either time dependence or time independence. MFLDA excels in both binary and multi-class classification problems. Its ability to identify key variables or signatures improves interpretability, which is important in biomedical research, as understanding these signatures can inform clinical decisions and future investigations. When applied to real-world data related to IBD, MFLDA outperformed existing methods. In addition, some of the selected variables are implicated in IBD,  while others could be explored for their potential role in IBD or related diseases.

Notable, MFLDA identified many pathways related to vitamin B deficiencies. Much of the existing literature predominantly focuses on single vitamin B deficiencies, and our approach highlights the importance of considering multiple vitamin B deficiencies in the context of IBD. Furthermore, we note that many studies use animal models primarily to investigate the implications of vitamin B deficiency in IBD \citep{selhub2013dietary}, indicating a need for more clinical studies to validate these findings in human populations. Currently, clinical trials \citep{ghishan2017vitamins} are exploring the potential of various vitamin B supplements to prevent the progression of IBD, which may further enrich our understanding of these pathways and their clinical relevance. Furthermore, vitamin B deficiencies can be linked to other diseases, such as cardiovascular disease \citep{albert2008effect}, neurodegenerative disorders \citep{kumar2022role}, and anemia \citep{morris2007folate}, suggesting that these conditions might share common pathways with IBD. This shared mechanistic understanding points to the potential for future research to analyze direct links between vitamin B deficiencies and these related diseases.

A main limitation of MFLDA is that it does not perform as well when only specific subdomains of time are discriminative between groups, as it assumes consistent discrimination across the entire time series. This limitation highlights an area for future work: the need to refine MFLDA or develop complementary methods to effectively identify and leverage these critical time subdomains. Despite this limitation, the simulation results and the findings of the real-data application are encouraging and could motivate several future applications and directions. One could focus on integrating multivariate longitudinal data with other types of data to enhance analysis. Furthermore, extending multivariate functional data to handle two-dimensional data, such as images, could broaden its applicability and provide even more comprehensive insight into biomedical research.


\section*{Funding and Acknowledgments}
Sandra E. Safo was partially supported by grant \#1R35GM142695 of the National Institute of General Medical Sciences of the National Institutes of Health. GuanNan Wang's research is partially supported by Simons Foundation Grant \#963447. The content is solely the responsibility of the authors and does not represent the official views of the funding agencies.

\section*{Data Availability and Software}
IBD data can be downloaded from The Inflammatory Bowel Disease Multiomics Database (\url{https://ibdmdb.org/}). The MFLDA package for implementing the method proposed, simulations and IBD data application is found at \url{https://github.com/liulim-Liu/MFLDA}. All statistical analyses are performed using R 4.3.3. 

\section*{Supplementary Material}
Web Tables and Figures referenced in Sections \ref{sec:simulation}, and \ref{sec:IBD} are available with this paper at the Biostatistics website. In the Supplementary Material, we also include hyperparameter $\tau$ tuning, algorithms, low ($n_i = 100, p = 100$) and high dimensional ($n_i = 500, p = 3000$ and $n_i = 1500, p = 1000$) binary simulation and results, additional multi-class simulation and results when features discriminating 3 classes are not identical. More information of IBD dataset can also be found including extraction, data processing steps, LMMs, binary problem (IBD vs. non-IBD) results, multi-class problem (CD vs. UC vd. non-IBD) and detailed interpretation of microbial pathways selected in IBD vs. non-IBD and CD vs. UC vs. non-IBD when assume time independence.

\clearpage
\bibliographystyle{unsrtnat}
\bibliography{References} 

\clearpage
\begin{center}
\textbf{\large Supplemental Materials to Multivariate Functional Linear Discriminant Analysis: An Application to Inflammatory Bowel Disease Classification}
\end{center}
\setcounter{equation}{0}
\setcounter{figure}{0}
\setcounter{table}{0}
\setcounter{page}{1}
\makeatletter
\renewcommand{\theequation}{S\arabic{equation}}
\renewcommand{\thefigure}{S\arabic{figure}}
\renewcommand{\bibnumfmt}[1]{[S#1]}
\renewcommand{\citenumfont}[1]{S#1}

\section{Hyperparameter Tuning}

The optimization in Equation (3) in the main article depends on the tuning parameter $\tau$ for $t$. Note that for $t$, $\tau > \|\mathcal{M}\widetilde{\bgamma}_1 \|_\infty$ results in a trivial solution vector, i.e., $\widehat{\bgamma}_1 = \mathbf{0}$. As such, we can set the upper bound for $\tau$ as $\tau_{\mathrm{max}} =  \|\mathcal{M}\widetilde{\bgamma}_1 \|_\infty$. For computational efficiency, instead of tuning $\tau$ for each time point $t_h$, we define an overall range that ensures the sparsity rate remains within $\pm5\%$ of a predetermined target (e.g., 10\% sparsity -- 10\% of features have nonzero coefficients). Specifically, we set the initial upper bound as $\tau_{max,0} = \max(\tau_{\mathrm{max}})$ and calculate the initial lower bound $\tau_{min,0} = \tau_{\mathrm{max}} \sqrt{\frac{\log p}{nT}}$. We then choose 8 even grid points between $(\tau_{min,0}, \tau_{max,0})$ and for each grid point we solve the optimization problem (Equation (2)) to obtain the discriminant scores at each time point. To determine whether a feature has non-zero coefficients across time points, we calculate the selectivity rate for that feature across all time points (e.g. 70\% selectivity implies for at least 70\% of time points the features have non-zero discriminant scores). If none of the $\tau$'s produces the desired sparsity rate, we update the initial $\tau$ range until we find a range of $\tau^{*}$ where the pre-specified sparsity rate is within the researcher's predefined range. \textbf{Algorithm 2} shows how we update to find the tuning range.

Once the range $(\tau_{\mathrm{min}}^*, \tau_{\mathrm{max}}^*)$, is identified, we use 5-fold cross-validation to select the optimal $\tau^{**}$ that maximizes a combined performance metric. This metric aggregates the accuracy (with range [0,1]), balanced accuracy (with range [0,1]), F-1 Score (with range [0,1]), precision (with range [0,1]), recall (with range [0,1]), and Matthew's Correlation Coefficient (MCC, with range [-1,1]), defined as:
$ \text{Accuracy} = \frac{\mathrm{TP}+\mathrm{TN}}{\mathrm{TP}+\mathrm{TN}+\mathrm{FP}+\mathrm{FN}}$;
$\text{Balanced Accuracy} = \frac{1}{2} (\frac{\mathrm{TP}}{\mathrm{P}} + \frac{\mathrm{TN}}{\mathrm{N}})$; 
$\text{F-1 Score} = \frac{\mathrm{TP}}{\mathrm{TP}+(\mathrm{FP}+\mathrm{FN})/2}$;
$\text{Precision} = \frac{\mathrm{TP}}{\mathrm{TP}+\mathrm{FP}}$;
$\text{Recall} = \frac{\mathrm{TP}}{\mathrm{TP}+\mathrm{FN}}$;
$\text{MCC} = \frac{\mathrm{TP} \times \mathrm{TN} - \mathrm{FP} \times \mathrm{FN}}{\sqrt{(\mathrm{TP}+\mathrm{FP})(\mathrm{TP}+\mathrm{FN})(\mathrm{TN}+\mathrm{FP})(\mathrm{TN}+\mathrm{FN})}}$.
Here, TP and TN stand for true positives and true negatives, respectively, and FP and FN stand for false positives/negatives. The higher the value of the criterion, the better the classification performance. The combined performance metric will be in the range [-1,6]. For a categorical result with more $K > 2$ classes, we use the weighted F-1 score, the weighted precision, and the weighted recall. The MCC then becomes \citep{gorodkin2004comparing, stoica2024pearson}: 
$\mathrm{MCC}=\frac{\sum_k \sum_l \sum_m C_{k k} C_{l m}-C_{k l} C_{m k}}{\sqrt{\sum_k\left(\sum_l C_{k l}\right)\left(\sum_{k^{\prime} \mid k^{\prime} \neq k} \sum_{l^{\prime}} C_{k^{\prime} l^{\prime}}\right)} \sqrt{\sum_k\left(\sum_l C_{l k}\right)\left(\sum_{k^{\prime} \mid k^{\prime} \neq k} \sum_{l^{\prime}} C_{l^{\prime} k^{\prime}}\right)}}$ for a $K \times K$ confusion matrix $C$. When there are more than two labels, the MCC will no longer be in the range $-1$ and $+1$. Instead, the minimum value will be between $-1$ and $0$ depending on the true "true positive" and "true negative". The maximum value is always $+1$. Thus, for multi-class problems, the combined performance metric will be in the range [-1,6] or [0,6]. We use the combined metric instead of a single metric for robustness. 

\section{Algorithm}
We use the \textbf{Non-Sparse Functional Linear Discriminant Analysis (LDA)} algorithm (i.e., Algorithm 1) to classify time-dependent functional data across multiple classes. We take as an input a functional data with $K$ classes, where each sample is a multivariate function. The algorithm computes the mean function for each class, using basis functions to represent the data. Then we calculate the between-class covariance matrix, which measures the separation between classes, and the within-class covariance matrix, which captures the variability within each class. Finally, the algorithm solves a generalized eigenvalue problem to find the projection function that maximizes class separability, outputting the non-sparse solution $\widetilde{\gamma}$.

{\setstretch{1.0}\begin{algorithm}[H]
\caption{Non-Sparse Functional LDA}
\label{alg:nonsparseLDA}
\begin{algorithmic}[1]
\State \textbf{Input:} $K$ classes of functional data: $\mathbf{Z_1} = \{\mathbf{z_1}^{(1)}(t), \mathbf{z_2}^{(1)}(t), \dots, \mathbf{z_{n_1}}^{(1)}(t)\}$, $\mathbf{Z_2} = \{\mathbf{z_1}^{(2)}(t), \mathbf{z_2}^{(2)}(t), \dots, \mathbf{z_{n_2}}^{(2)}(t)\}$, $\dots$, $\mathbf{Z_k} = \{\mathbf{z_1}^{(k)}(t), \mathbf{z_2}^{(k)}(t), \dots, \mathbf{z_{n_k}}^{(k)}(t)\}$ where $\mathbf{z_i}^{(k)}(t) \in \Re^p$ and $t \in T$ is a continuous domain.
\State \textbf{Output:} non-sparse solution $\widetilde{\mathbf{\gamma}}$.

\State \textbf{Step 1:} Compute the mean function of each class:
\begin{equation*}
    \Bar{z}_k(t) = \frac{1}{n_k} \sum_{i \in G_k} z_i^{(k)}(t) = \frac{1}{n_k} \sum_{i \in G_k} [\bphi^T(t)\mathbf{c}_{i1}^{(k)} ... \bphi^T(t)\mathbf{c}_{ip}^{(k)}] \bbeta(t) = [\bphi^T(t)\mathbf{\Bar{c}}_{1}^{(k)} ... \bphi^T(t)\mathbf{\Bar{c}}_{p}^{(k)}]\bbeta(t)
\end{equation*}

\State \textbf{Step 2:} Compute the between-class covariance matrix $\bSb(t)$:
\begin{equation*}
    \bSb = \sum_{k = 1}^G n_k \mathbf{A}_k(t) = \begin{bmatrix} 
    (\mathbf{\bar{c}}_{1}^{(k)} - \mathbf{\bar{c}}_{1})^T\\
    \vdots \\
    (\mathbf{\bar{c}}_{p}^{(k)} - \mathbf{\bar{c}}_{p})^T 
\end{bmatrix} 
\bPhi(t) \bPhi(t)^T 
[(\mathbf{\bar{c}}_{1}^{(k)} - \mathbf{\bar{c}}_{1}) ...  (\mathbf{\bar{c}}_{p}^{(k)} - \mathbf{\bar{c}}_{p})]
\end{equation*}

\State \textbf{Step 3:} Compute the pooled within-class covariance matrix $\mathbf{S}_p (t_h)$ for each time point $t_h$ independently:
\begin{equation*}
    \mathbf{S}_p = \frac{1}{\sum_{k = 1}^G (n_k-1)} \left(\sum_{k = 1}^G (n_k -1) \mathbf{S}^{(k)}\right)
\end{equation*}
where the $jq$th entry in $\bS^{(k)}(t_h,t_s)$ is 
\begin{equation*}
    s_{jq}^{(k)}(t_h, t_s) 
= \frac{1}{n_k - 1} \sum_{i \in G_k}
\Bigl(c_{ij}^{(k)} - \bar{\mathbf{c}_j}^{(k)}\Bigr)^{\smt} \bphi(t_h) \bphi(t_s)^{\smt} \Bigl(c_{iq}^{(k)} - \bar{\mathbf{c}_q}^{(k)}\Bigr).
\end{equation*}

\State \textbf{Step 4:} Compute the optimal projection function $\bbeta(t)$ by solving the functional generalized eigenvalue problem:
\begin{equation*}
    \max_{\bbeta} \bbeta^T \bSb \bbeta ~~~\mbox{subject to}~~ \bbeta^T \bSp \bbeta = 1   
\end{equation*}

Let $\mathcal{M}=\bS_p^{{-1/2}}\bS_{b}\bS_p^{{-1/2}}$ and $(\widetilde{\lambda}, \widetilde{\gamma})$ be its eigenvalue-vector pair, which are solutions to the eigenvalue system:
\begin{equation*}
\mathcal{M}\bgamma= \lambda\bgamma
\end{equation*}

\State \textbf{Step 5:} Choose the eigenfunction corresponding to the largest eigenvalue as the projection function $\widetilde{\bgamma}$.

\State \textbf{Step 6:} Without loss of generality, assume that data is time independent and calculate $\widetilde{\bgamma}$ for each time points.

\end{algorithmic}
\end{algorithm}}

We use the \textbf{Tuning Range} algorithm to determine an optimal tuning range \([\tau_{\mathrm{min}}, \tau_{\mathrm{max}}]\) for a data tensor \(\bX\) given a predefined sparsity rate \(\rho\). It starts by solving an optimization problem to obtain a non-sparse discriminant vector. Using this, it calculates an initial tuning range by determining an upper bound \(\tau_{\mathrm{max}}\) and a corresponding lower bound \(\tau_{\mathrm{min}}\). The algorithm then evaluates the sparsity of the features across a grid of \(\tau\) values within this range. Based on the observed sparsity, it iteratively updates the tuning range until the sparsity matches the expected rate. The final optimal range is then output as the tuning range.

{\setstretch{1.0}\begin{algorithm}[H]
\caption{Tuning Range $\tau$}
\label{alg:TuningRange}
\begin{algorithmic}[1]
    \State \textbf{Input:} Data tensor $\bX \in \Re^{n \times p \times T}$, predefined sparsity rate $\rho$.
    \State \textbf{Output:} Tuning range $[\tau_{\mathrm{min}}, \tau_{\mathrm{max}}]$

    \State \textbf{Step 1:} For time point $t$, solve the optimization problem using Algorithm \ref{alg:nonsparseLDA} to obtain the non-sparse discriminant vector \(\widetilde{\bgamma}(t)\).

    \State \textbf{Step 2:} Obtain Initial Tuning Range \(\tau\):
    \begin{enumerate}
        \item Calculate the initial upper bound: \(\tau_{\mathrm{max}} = \max\left(\|\mathcal{M}(t)\widetilde{\bgamma}_1(t) \|_\infty\right)\).
        \item Calculate the initial lower bound: \(\tau_{\mathrm{min}} = \tau_{\mathrm{max}} \sqrt{\frac{\log p}{nT}}\).
        \item Select 8 grid points evenly spaced between \(\tau_{\mathrm{min}}\) and \(\tau_{\mathrm{max}}\).
    \end{enumerate}

    \State \textbf{Step 3:} Determine Sparsity:
    \begin{enumerate}
        \item For each \(\tau\) in the grid, calculate the discriminant scores for each feature at each time point.
        \item Calculate the selectivity rate (percentage of time points with non-zero discriminant scores) for each feature.
        \item Determine the sparsity (percentage of features with selectivity rates within the expected range).
    \end{enumerate}

    \State \textbf{Step 4:} Update Tuning Range:
    \begin{enumerate}
        \item If the sparsity is within the expected range, set the current \(\tau\) as the new upper bound and the previous \(\tau\) as the new lower bound.
        \item If the sparsity is less than expected, decrease \(\tau\) by dividing by a constant \(c\), and repeat until within the expected range.
        \item If the sparsity is greater than expected, increase \(\tau\) by multiplying by a constant \(c\), and repeat until within the expected range.
        \item Once the optimal range \((\tau_{\mathrm{min}}^*, \tau_{\mathrm{max}}^*)\) is identified, select 8 grid points within this range.
    \end{enumerate}

    \State \textbf{Step 5:} Output the final tuning range $[\tau_{\mathrm{min}}, \tau_{\mathrm{max}}] = $ \((\tau_{\mathrm{min}}^*, \tau_{\mathrm{max}}^*)\)
\end{algorithmic}
\end{algorithm}}

We use the \textbf{Cross-Validation algorithm with convergence for Mulrivariate Functional Linear Discriminant Analysis (MFLDA)} identify the optimal regularization parameter, \(\tau\), to enhance classification accuracy in time-dependent data. MFLDA is particularly suited for high-dimensional, temporal datasets, enabling both classification and feature selection. The algorithm starts by taking in a data tensor and a range of \(\tau\) values, splitting the data into \(K\) stratified folds. For each fold, the model is trained on \(K-1\) folds and validated on the remaining one, iteratively solving the MFLDA optimization problem for each \(\tau\) until convergence. Discriminant vectors are computed and used to classify the validation set, with performance metrics such as accuracy, precision, and recall evaluated. After averaging the results across all folds, the \(\tau\) that maximizes the combined performance metric is selected. Finally, the model is refitted on the full dataset using the optimal \(\tau\), and the final discriminant vector, classification results, and feature selection matrix are reported.

{\setstretch{1.0}
\begin{algorithm}[H]
\caption{Cross-Validation for Choosing \(\tau\) and Solving Multivariate Functional LDA (MFLDA) until Convergence}
\label{alg:CV_tau_MFLDA}
\begin{algorithmic}[1]
    \State \textbf{Input:} Data tensor \(\bX \in \Re^{n_i \times p \times T_{i}}\) where $T_{i}$ represent number of time input for each individual $i$, range of \(\tau\) values \(\{\tau_v\}_{v=1}^8\), number of folds \(K = 5\).
    \State \textbf{Output:} Optimal tuning parameter \(\tau^{**}\), final discriminant vector \(\widehat{\bgamma}\).

    \State \textbf{Step 1:} Initialize:
    \begin{enumerate}
        \item Complete data: estimate the complete data curve from irregular time using b-spline estimation, having data tensor \(\bX \in \Re^{n \times p \times T}\) in the same
time domain for each feature.
        \item Standardize data: center and scale \(\bX\) by time point and by feature.
        \item Split the data into \(K\) folds using class stratification to ensure proportional class representation in each fold.
    \end{enumerate}

    \State \textbf{Step 2:} For each fold \(k = 1,\ldots,K\):
    \begin{enumerate}
        \item Use \(k^{th}\) fold as the validation set and the remaining \(K-1\) folds as the training set.
        \item \textbf{For each \(\tau_v\) in the range \(\{\tau_v\}_{v=1}^8\) obtained using \textbf{Algorithm \ref{alg:TuningRange}}:}
        \begin{enumerate}
            \item Compute the initial discriminant vector \(\widehat{\bgamma}_1^{\tau_v}\) by solving the optimization problem in \textbf{Algorithm \ref{alg:nonsparseLDA}}.
            \item Repeat the optimization until convergence to obtain the final discriminant vector \(\widehat{\bgamma}_1^{\tau_v}\) for all time point \(t\).
            \item Project the training data onto \(\widehat{\bgamma}_1^{\tau_v}\) to compute discriminant scores.
            \item Classify the validation data by projecting it onto \(\widehat{\bgamma}_1^{\tau_v}\) and applying the nearest centroid algorithm. If the data are time-dependent, classification is performed for each time point; if time-independent, classification is done overall.
            \item Calculate performance metrics (accuracy, balanced accuracy, F-1 score, precision, recall, MCC) for the \(k^{th}\) fold.
        \end{enumerate}
    \end{enumerate}

    \State \textbf{Step 3:} Select Optimal \(\tau^{**}\):
    \begin{enumerate}
        \item For each \(\tau_v\), average the performance metrics across all folds.
        \item Choose \(\tau^{**}\) that maximizes the combined performance metric (sum of accuracy, balanced accuracy, F-1 score, precision, recall, and MCC).
    \end{enumerate}

    \State \textbf{Step 4:} Final Model:
    \begin{enumerate}
        \item Use the full dataset to solve MFLDA with \(\tau^{**}\) until convergence to obtain \(\widehat{\bgamma}^{\tau^{**}}\).
        \item Classify test data using \(\widehat{\bgamma}^{\tau^{**}}\) and compute the final performance metrics.
        \item Generate visualizations of discriminant scores and classification results.
        \item Test the similarity of class assignment using the Kolmogorov-Smirnov test.
        \item Output the final variable selection matrix \(\widehat{\bgamma}^{\tau^{**}}\) for all time points.
    \end{enumerate}
\end{algorithmic}
\end{algorithm}
}

\section{Simulation selected and non selected features}

\begin{figure}[H]
    \centering
        \includegraphics[width=\textwidth]{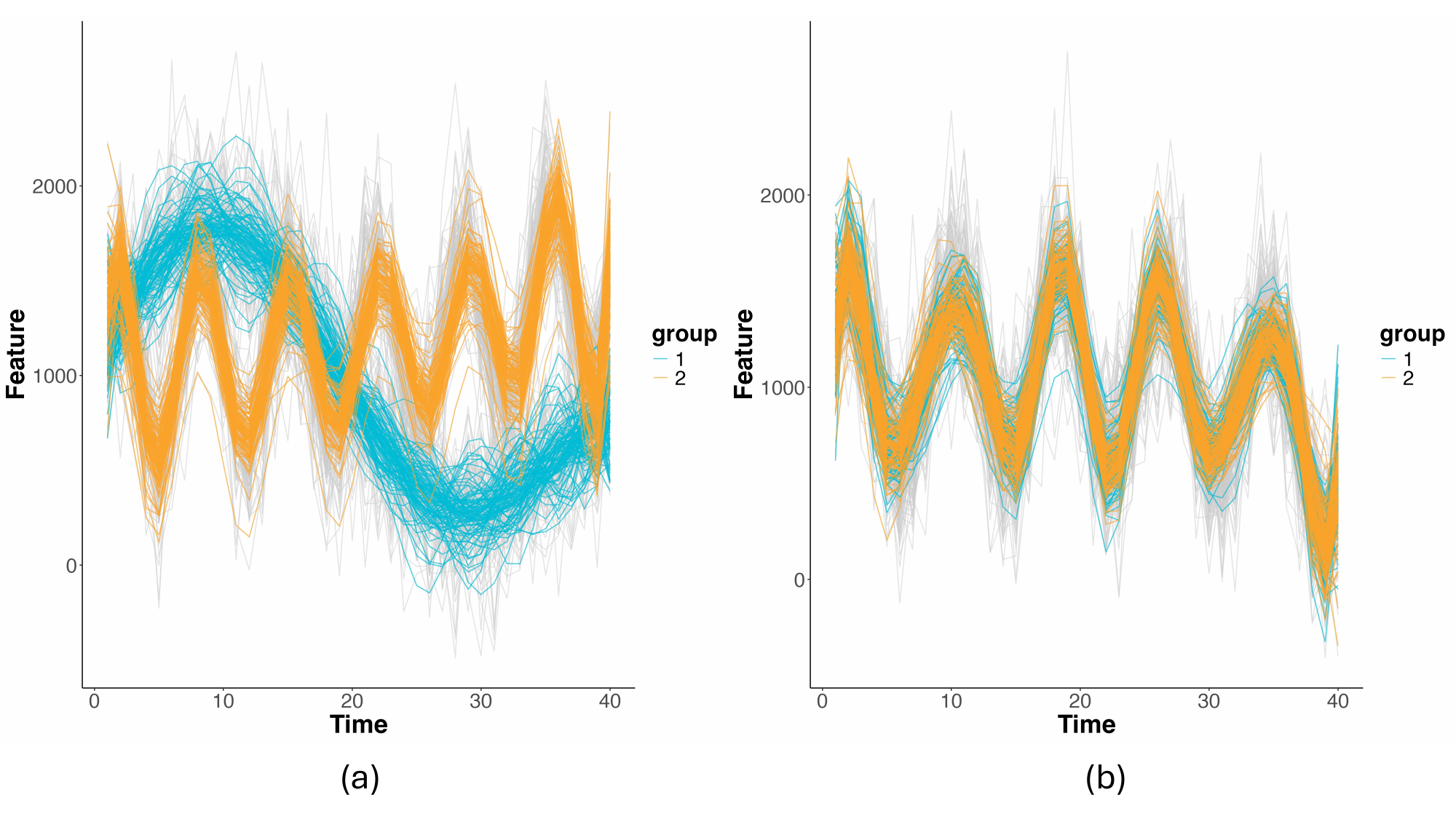}
        \caption{Selected Variables and Not Selected Variables. Gray lines represent raw simulation and colored lines represent b-spline smoothed estimation colored by groups: (a) discriminant variables; (b) non-discriminant variables}
        \label{fig:varselected}
\end{figure}

In all simulation studies, we define selected and non-selected features. Figure \ref{fig:varselected} shows one feature that can distinguish between groups, and one feature that cannot distinguish between groups. In Figure \ref{fig:varselected}(a), the two groups are different in times with similar mean and different patterns. But in Figure \ref{fig:varselected}(b), we cannot distinguish the two groups in time. 

\section{Binary-Class Simulation}
We consider a binary classification problem ($K = 2$) with low [($n_k=100, p=100$)] and high [($n_k=500, p=3000$), ($n_k=1500, p=1000$)] dimensional settings and assume that few variables are signals and differ between classes, while other variables have little or no differences between the groups. In all examples, we simulate the data to have $T=40$ time points and consider different mean separation functions and time periods where the groups differ. We simulate 100 Monte Carlo replicates and report average performance metrics.
We generate a set of curves over the time interval $T = [1,40]$ for $n_1$ samples belonging to group 1 and $n_2$ samples in group 2. We generate $p$ variables of function curves for each sample. Following \citep{gardner2021linear}, each set of functions was generated without within-subject temporal structure by: $$x_{ij}(t) = \lambda \times \delta + \eta_{0j} + \eta_{1j}t + \eta_{2j}t^2 + \eta_{3j}t^3 + \eta_{4j}t^4 + \eta_{5j}\sin(\eta_{6j}t) + \epsilon_{ij},$$ where $\eta_{0j}, \eta_{1j}, \eta_{2j}, \eta_{3j}, \eta_{4j}, j = 1,...,p$, are generated from the least squared fit of a fourth-degree polynomial to $(x_{1j}^{*}, y_{1j}^{*}), (x_{2j}^{*}, y_{2j}^{*}), (x_{3j}^{*}, y_{3j}^{*}), (x_{4j}^{*}, y_{4j}^{*}), (x_{5j}^{*}, y_{5j}^{*}), (x_{6j}^{*}, y_{6j}^{*})$. Here $x_1^* = 0$, $x_6^* = 10, x_k^* \sim \text{unif}(0,10), k = 2, 3, 4, 5$ and $y_k^* \sim \text{unif}(50,100), k = 1,\ldots,6$. $\eta_{5j} = \max(\eta_{0j} + \eta_{1j}t + \eta_{2j}t^2 + \eta_{3j}t^3 + \eta_{4j}t^4) - \min (\eta_{0j} + \eta_{1j}t + \eta_{2j}t^2 + \eta_{3j}t^3 + \eta_{4j}t^4)$ under the interval $t = 1, 2, \ldots, 40$, and $\eta_{6j} \sim \text{unif}(0,10)$, $\epsilon_{p,ij} \sim N(0, \sigma^2)$. The value of $\delta$ represents the separation between the two groups, $\delta = 0$ for group 1, and $\delta = 500$ for group 2. The indicator $\lambda$ represents which group has a higher mean function, $\lambda = 1$ for group 1, and $\lambda$ is randomly chosen from 1 or -1 for group 2. 

\begin{figure}[h]
    \centering
        \includegraphics[width=\textwidth]{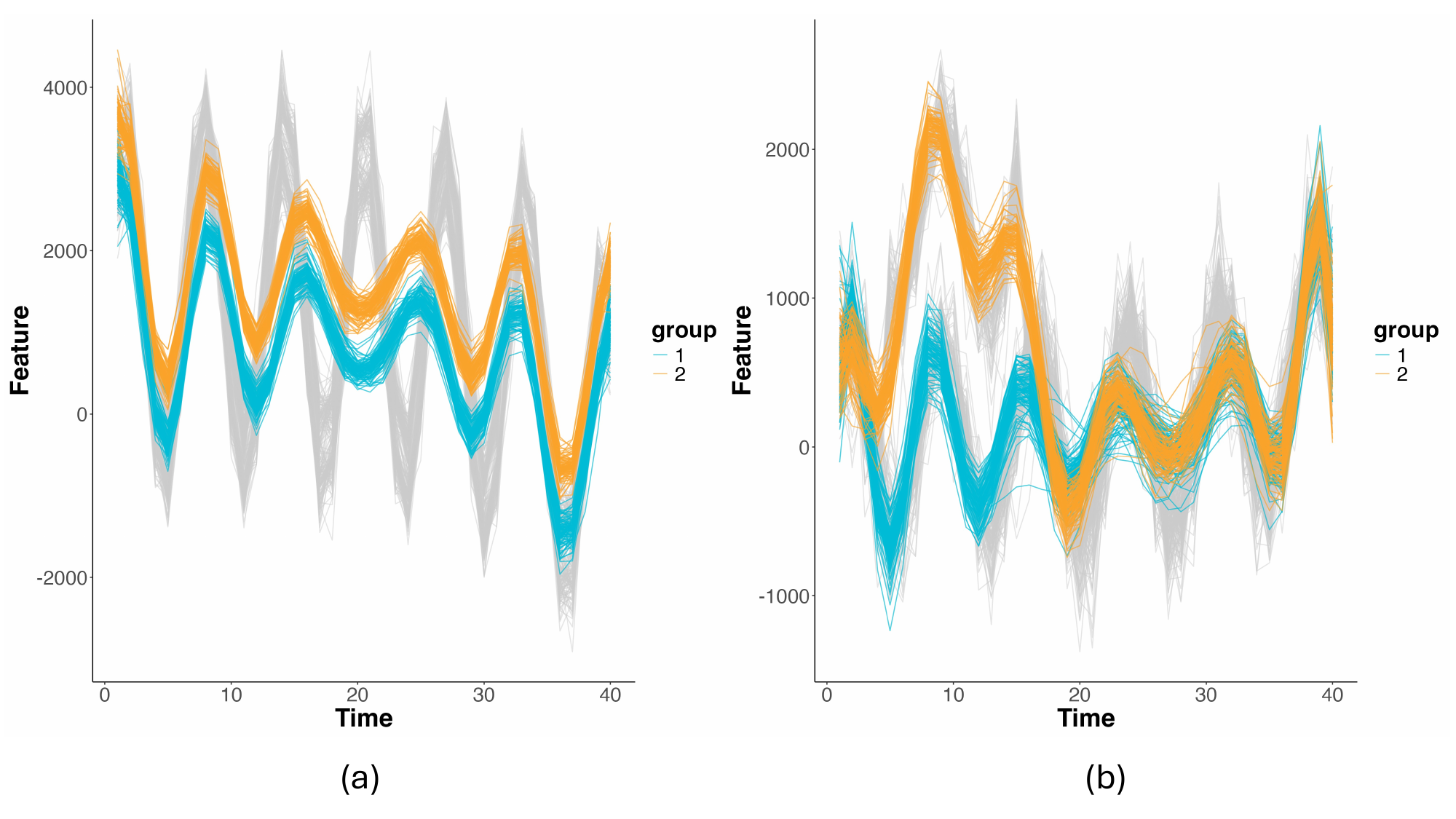}
        \caption{Illustration of Simulation Datasets. Gray line represents the raw simulation and colored lines are B-spline smoothed estimations for each group: (a) groups separated at all time points; (b) groups only separated between [5,15].}
        \label{fig:simulation}
\end{figure}

We generate multiple simulation data where the first ten features have a greater separation between two groups, and the other 90 features have little or no differences between the two groups. We consider multiple simulation cases and generate data as follows: (1) two groups are separated at all time points with similar mean functions; (2) two groups are separated only between a fixed time period [5,15] with different patterns; (3) two groups are only separated between a period of 10 time points with different patterns. (4) Two groups are only separated between a random time period (random length between 5 and 40) with different mean functions and patterns.  Figure \ref{fig:simulation} is a pictorial representation of two cases where the gray lines represent true observed data, and the colored lines represent smoothed B-spline estimations. As shown in Figure \ref{fig:simulation} (a), the colors distinguish the groups, which exhibit different means across all time points, although their overall patterns remain consistent. In contrast, in some real-world applications, groups differ only at specific time points rather than at all time points. In the second case, Figure \ref{fig:simulation} (b) shows the two groups exhibit the same mean function except during the time period [5, 15]. Group differences do not always occur within the same time period; thus, in case (3), the groups differ over the same length of time (10 time points) but across random time intervals. In case (4), group differences occur at random time periods, with varying lengths of divergence. Group differences may manifest in distinct mean functions or involve similar mean functions but with different patterns.

\subsection{Binary Simulation Results and Comparison}

\begin{table}[h]
    \caption{Binary Classification and feature selection test metrics for simulated data based on 100 repetitions. The highest metrics are in bold.}
    \label{tab:simresults}
    \centering
    \begin{tabular}{c|c|ccc|ccc}
    \hline
    \hline
    & & \multicolumn{3}{c|}{Classification} & \multicolumn{3}{c}{Feature Selection} \\ \cline{3-8}
    Case & Model & Sens & Spec & F-1 & Sens & Spec & F-1\\
    \hline
    Different in & MFLDA-I   &  \textbf{1.00} & \textbf{1.00}  & \textbf{1.00} & \textbf{1.00} & 0.90 & \textbf{0.94} \\
    all [0, 40] & PLDA   &  \textbf{1.00} & \textbf{1.00}  & \textbf{1.00}  & \textbf{1.00}  & 0.52 & 0.66 \\
    & FLDA Best & \textbf{1.00} & 0.98 & 0.99 & - & - & -\\
    & FPCA+PLDA & 0.94 & 0.97 & 0.96 & 0.80 & \textbf{1.00} & 0.89 \\
    \hline
    Different only & MFLDA-I   &  \textbf{0.99} & 0.98  &  \textbf{0.98} & \textbf{0.98} & 0.50 & 0.63 \\
    in [5, 15] & PLDA   &  0.52 & 0.76  &  0.68 & 0.50 & 0.61 & 0.31 \\
    & FLDA Best & 0.86  & \textbf{1.00}  &  0.93 & - & - & -\\
    & FPCA+PLDA &  0.96 &  0.96 & 0.97  & 0.70  & \textbf{0.99} & 0.78 \\
    \hline
    Different in &   MFLDA-I   &  \textbf{1.00} & \textbf{1.00}  &  \textbf{1.00} & \textbf{0.99} & \textbf{1.00}  & \textbf{0.95}\\
    random 10 &   PLDA   & 0.97  &  0.98 & 0.98  & 0.95 & 0.30 & 0.38 \\
    time point &   FLDA Best   & \textbf{1.00}  & 0.88  &  0.94 & - & - & -\\
       &   FPCA+PLDA   &  \textbf{1.00} & \textbf{1.00}  &  \textbf{1.00} &  0.80 & \textbf{1.00}& 0.89 \\
    \hline
    Different in   &   MFLDA-I   &  \textbf{1.00} & \textbf{1.00}  &  \textbf{1.00} & \textbf{1.00} & \textbf{1.00} & \textbf{1.00} \\
    random time   &   PLDA   & 0.99  & 0.99  &  0.99 & 0.97 & 0.29 & 0.40\\
    and random   &   FLDA Best   &  \textbf{1.00} &  0.92 &  0.96 & - & - & -\\
    length of time   &   FPCA+PLDA   & \textbf{1.00}  &  \textbf{1.00} &  \textbf{1.00}  &  0.90 & 0.98 &  0.90 \\
    \hline
    \hline
    \end{tabular}

\end{table}

\begin{figure}[h!]
    \centering
        \includegraphics[width=\textwidth]{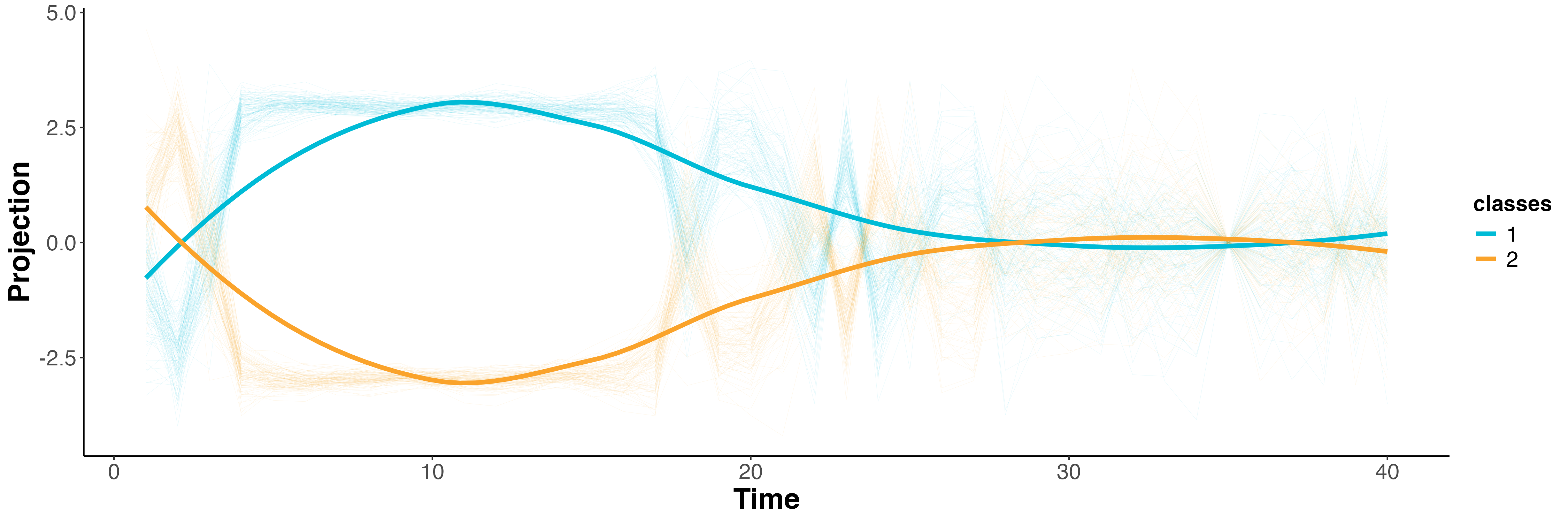}
        \caption{Discriminant Plot of Case 2. Thin lines represented the projection for each subjects. The think lines represented the LOESS curves for each class colored by predicted class of testing case subjects.}
        \label{fig:discriminant}
\end{figure}

In this simulation setting, we assume variables are time independent (no correlations between time points). We compare the performance of proposed and existing methods on simulated test data using metrics that include sensitivity, specificity, and F1 score. The test data set uses the same distribution as the training data and have the the same dimensions. Table \ref{tab:simresults} displays the classification and variable performance of the proposed method in comparison with existing methods.  Our proposed MFLDA-I method outperforms the other three methods in all cases. Figure \ref{fig:discriminant} shows a discriminant plot for better visualization. In the discriminant plot, a smoothed LOESS curve is produced. We observe that the two groups are well-separated during the time period [5, 23], and more so between time periods [5,15], and after time 23, the two groups are not distinguishable. This finding is consistent with case 2. The Kolmogorov-Smirnov test also provides a significant conclusion of a p-value less than 0.05, suggesting that the two curves are different.

MFLDA-I can also find key variables that distinguish between two classes. In comparison, the LDA for multiple functional data (FLDA Best) by Gardner-Lubbe \citep{gardner2021linear} was not developed to select features. Thus for feature selection, we compare our method with PLDA and a two-step method that combines FPCA and PLDA (refer to Table \ref{tab:simresults}). In most cases, our methods consistently find most of the relevant variables with high F-1 score, sensitivity, and specificity values, which are sometimes better than competing methods. However, in Case 2 where the classes discriminate with variables that are only different from time 5 to time 15, we obtain moderate an F-1 score and sensitivity values In this case, PLDA also has smaller sensitivity and specificity values than the other three cases, but FPCA + PLDA outperformed all methods in terms of specificity and F-1 score. In the other three cases, PLDA has high sensitivity and low specificity, meaning it is less likely to miss important variables (fewer false negatives). However, the trade-off is that its low specificity leads to a higher rate of false positives, resulting in more irrelevant variables being selected.
The performance of FPCA + PLDA is comparable to our proposed MFLDA-I method in many instances, with occasional cases where FPCA + PLDA achieves slightly better results. However, our MFLDA-I method offers several key advantages. Unlike FPCA + PLDA, which separates feature extraction and classification into two steps, MFLDA-I integrates classification and feature selection simultaneously. This integration not only simplifies the workflow but also allows for more interpretable variable selection, as it directly identifies the most relevant features contributing to the classification. Additionally, our method avoids potential information loss or distortion that may occur during the FPCA dimensionality reduction step, ensuring a more direct and efficient utilization of the original data.

\section{Binary-Class High-Dimensional Simulation}
In this section, we present the results from a large-scale simulation study designed to evaluate the performance of our method under different data regimes. We extend the evaluation to two additional scenarios: \(n \gg p\) and \(p \gg n\). These cases are crucial for understanding how the method scales and adapts to datasets where the number of observations vastly exceeds the number of features (\(n \gg p\)) and vice versa (\(p \gg n\)), providing insights into its robustness and applicability across different practical settings. For computational efficiency, we assume variables are time independent in high-dimensional simulation.

\subsection{$N_i = 500, P = 3000$}

\begin{table}[h]
    \caption{High-Dimensional($N_i = 500, P = 3000$) Binary Classification (left) and Variable Selection (right) test metrics for binary simulated data based on 100 repetitions. The highest metrics are in bold.}
    \label{tab:large1-results}
    \centering
    \begin{tabular}{c|c|ccc|ccc}
    \hline
    \hline
         &       & \multicolumn{3}{c|}{Classification} & \multicolumn{3}{c}{Feature Selection} \\ \cline{3-8}
         Case &  Model & Sens & Spec & F-1 & Sens & Spec & F-1\\
    \hline
    Different in   &   MFLDA-I   & \textbf{1.00} & \textbf{1.00} & \textbf{1.00} & 0.68 & \textbf{1.00} & \textbf{0.81} \\
    all [0,40]   &   PLDA   & \textbf{1.00}  & \textbf{1.00} & \textbf{1.00} & \textbf{1.00} & 0.00 & 0.18\\
    & FLDA Best & - & - & -& - & - & -\\
       &   FPCA+PLDA   & \textbf{1.00}  & \textbf{1.00} & \textbf{1.00} & \textbf{1.00} & 0.00 & 0.18\\
    \hline
    Different only   &   MFLDA-I   & \textbf{1.00}  & \textbf{1.00} & \textbf{1.00} & 0.93 & \textbf{0.94} & \textbf{0.96}\\
    in [5,15]   &   PLDA   & 0.95  & 0.94 & 0.95 & \textbf{1.00} & 0.00 & 0.18\\
    & FLDA Best & - & - & -& - & - & -\\
       &   FPCA+PLDA   & \textbf{1.00}  & \textbf{1.00} & \textbf{1.00} & \textbf{1.00} & 0.00 & 0.18\\
    \hline
    Different in random   &   MFLDA-I   & \textbf{1.00}  & \textbf{1.00} & \textbf{1.00} & 0.77 & \textbf{0.99} & \textbf{0.87}\\
    10 time points   &   PLDA   & \textbf{1.00}  & \textbf{1.00} & \textbf{1.00} & 0.95 & 0.05 & 0.18\\
    & FLDA Best & - & - & -& - & - & -\\
       &   FPCA+PLDA   & \textbf{1.00}  & \textbf{1.00} & \textbf{1.00} & \textbf{1.00} & 0.00 & 0.18\\
    \hline
    Different in random time   &   MFLDA-I   & \textbf{1.00}  & \textbf{1.00} & \textbf{1.00} & 0.79 & \textbf{0.99} & \textbf{0.88}\\
    and random length of time  &   PLDA   & \textbf{1.00}  & \textbf{1.00} & \textbf{1.00} & 0.98 & 0.03 & 0.18\\
    & FLDA Best & - & - & -& - & - & - \\
       &   FPCA+PLDA   & \textbf{1.00}  & \textbf{1.00} & \textbf{1.00} & \textbf{1.00} & 0.00 & 0.18\\
    \hline
    \hline
    \end{tabular}

\end{table}

Table \ref{tab:large1-results} summarizes the performance of our MFLDA-I method compared to PLDA, FLDA Best, and FPCA + PLDA in different simulation scenarios. Across all scenarios, MFLDA-I consistently achieved perfect classification performance, with Sensitivity, Specificity, and F-1 scores of 1.00. This indicates that MFLDA-I is highly effective in distinguishing between classes, irrespective of the underlying data structure. Meanwhile,  FLDA Best cannot produce the results due to computation inefficiency.

In terms of feature selection, MFLDA-I demonstrated superior performance compared to the other methods across  all scenarios, especially for specificity and F-score. For instance, in the ``Different in [5,15]'' scenario, MFLDA-I, had a higher Specificity (0.94) and a higher F-1 score (0.96) compared to the other methods. PLDA and FPCA+PLDA also have perfect classification performance but suboptimal variable selection as evidenced by its lower specificty and F-1 values.

The robustness of MFLDA-I is further highlighted in more complex scenarios such as "Different in random time" and "Different in random length of time," where MFLDA-I maintained high variable selection performance with F-1 scores of 0.87 and 0.88, respectively. In contrast, the competing methods struggled to accurately identify the relevant variables, with significantly lower specificity and F-1 scores.

\subsection{$N_i = 1500, P = 1000$}

\begin{table}[h]
    \caption{High-Dimensional($N_i = 1500, P = 1000$) Binary Classification (left) and Variable Selection (right) test metrics for binary simulated data based on 100 repetitions. The highest metrics are in bold.}
    \label{tab:large2-results}
    \centering
    \begin{tabular}{c|c|ccc|ccc}
    \hline
    \hline
         &       & \multicolumn{3}{c|}{Classification} & \multicolumn{3}{c}{Feature Selection} \\ \cline{3-8}
    Case &  Model & Sens & Spec & F-1 & Sens & Spec & F-1\\
    \hline
    Different in   &   MFLDA-I   & \textbf{1.00} & \textbf{1.00} & \textbf{1.00} & \textbf{1.00} & \textbf{1.00} & \textbf{1.00}\\
    all [0,40]   &   PLDA   & \textbf{1.00} & \textbf{1.00} & \textbf{1.00} & \textbf{1.00} & 0 & 0.18\\
    & FLDA Best & - & - & -& - & - & - \\
       &   FPCA+PLDA & \textbf{1.00} & \textbf{1.00} & \textbf{1.00} & \textbf{1.00} & 0 & 0.18\\
    \hline
    Different only   &   MFLDA-I   & \textbf{1.00} & \textbf{1.00} & \textbf{1.00} & 0.94 & \textbf{0.86} & \textbf{0.96}\\
    in [5,15]   &   PLDA   & 0.94 & 0.94 & 0.94 & \textbf{1.00} & 0.00 & 0.18 \\
    & FLDA Best & - & - & -& - & - & -\\
       &   FPCA+PLDA & \textbf{1.00} & \textbf{1.00} & \textbf{1.00} & \textbf{1.00} & 0.00 & 0.18\\
    \hline
    Different in random   &   MFLDA-I & \textbf{1.00} & \textbf{1.00} & \textbf{1.00} & 0.93 & \textbf{0.98} & \textbf{0.96}\\
    10 time points   &   PLDA   & \textbf{1.00} & \textbf{1.00} & \textbf{1.00} & 0.74  & 0.28 & 0.16\\
    & FLDA Best & - & - & -& - & - & -\\
       &   FPCA+PLDA & \textbf{1.00} & \textbf{1.00} & \textbf{1.00} & \textbf{1.00} & 0.00 & 0.18\\
    \hline
    Different in random time   &   MFLDA-I & \textbf{1.00} & \textbf{1.00} & \textbf{1.00} & 0.93 & \textbf{0.95} & \textbf{0.96}\\
    and random length of time  &   PLDA   & \textbf{1.00} & \textbf{1.00} & \textbf{1.00} & 0.81 & 0.20 & 0.16\\
    & FLDA Best & - & - & -& - & - & -\\
       &   FPCA+PLDA & \textbf{1.00} & \textbf{1.00} & \textbf{1.00} & \textbf{1.00} & 0.00 & 0.18\\
    \hline
    \hline
    \end{tabular}

\end{table}

Table \ref{tab:large2-results} presents the performance comparison of MFLDA-I against PLDA, FLDA Best and FPCA + PLDA , evaluating both classification accuracy and the effectiveness of feature selection in different simulation settings. In terms of classification, all methods consistently achieve a perfect score in all scenarios, with Sensitivity, Specificity, and F-1 scores all reaching 1.00. This uniformity in performance highlights the reliability of MFLDA-I in accurately identifying class distinctions, even under varying conditions. While FLDA Best cannot produce the results due to computation inefficiency. When examining feature selection, MFLDA-I shows similar performance to Section 5.1.


Furthermore, in the "Different in random 10 time points" and "Different in random time and random length of time" scenarios, MFLDA-I continued to excel in feature selection, evidenced by F-1 scores of 0.96. Penalized LDA, though it also performed well in classification, showed weaker results in feature selection, with significantly lower Specificity and F-1 scores compared to MFLDA-I.

Overall, these results demonstrate that MFLDA-I not only excels in classification accuracy but also significantly outperforms other methods in identifying important features, particularly in complex and high-dimensional datasets. This makes MFLDA-I a highly effective method for both classification and feature selection in diverse data scenarios.


\section{More on Multi-Class Simulation}
Another possible scenario could be that group differences occur in different features: features differentiating group 1 and group 2 are different from features differentiating group 2 and group 3.

\begin{figure}[h]
    \centering
        \includegraphics[width=\textwidth]{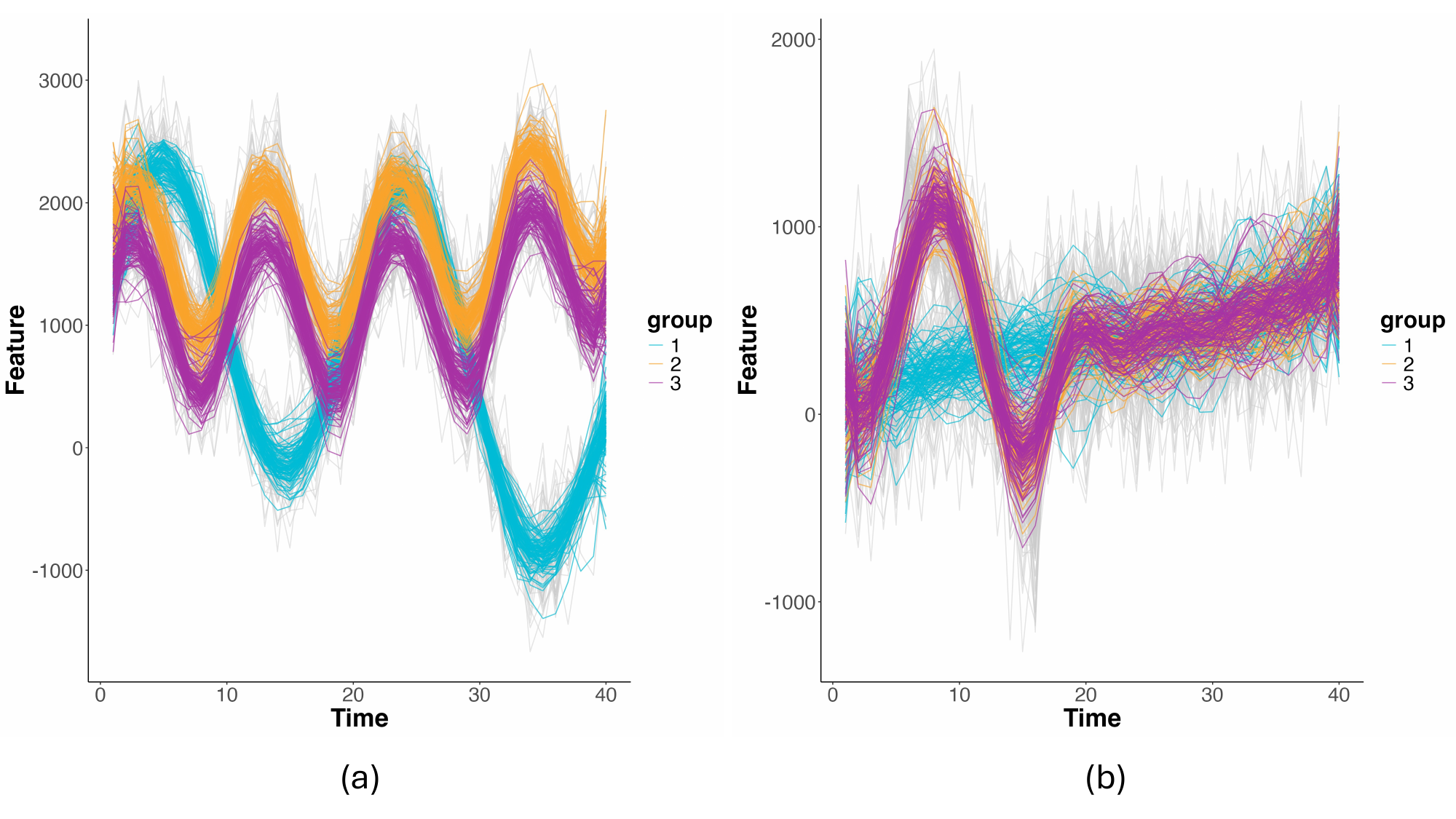}
        \caption{Illustration of Multi-Class Simulation Datasets. Gray lines represent raw simulation and colored lines represent b-spline smoothed estimation colored by groups: (a) groups separated at all time points; (b) variables for distinguishing between group 1 and group 2 with separation between [5,15]}
        \label{fig:simulation-3}
\end{figure}

In this scenario we generated a set of curves over the time interval $T = [1,40]$ which contains $n_1$ samples for group 1, $n_2$ samples for group 2 and $n_3$ samples for group 3 with $p$ features for each sample. Only the first 10\% of the features can distinguish between group 1 and group 2, and other different 10\% of features can distinguish between group 2 and group 3. The set of functions are generated without within-subject temporal structure as follows: $x_{ij}(t) = \lambda \times \delta + \eta_{0j} + \eta_{1j}t + \eta_{2j}t^2 + \eta_{3j} t^3 + \eta_{4j} t^4 + \eta_{5j} \sin(\eta_{6j}t) + \epsilon_{ij}$, for feature $j$, if the feature is been selected to discriminate between group 1 and group 2, then the value of $\delta$ represents the separation between the three groups, $\delta = 0$ for group 1, and $\delta = 500$ for group 2 and group 3. Similarly, for features that separate group 2 and 3, we have $\delta = 0$ for group 2, and $\delta = 500$ for group 3 and group 1. The indicator $\lambda$ represents which group has a higher mean function, $\lambda = 1$ for one group, and $\lambda$ is randomly chosen from 1 or -1 for the other two groups. In the simulation, 2 of the groups have similar patterns and similar mean functions across all time or partial time. For example, Figure \ref{fig:simulation-3}(a) shows group 2 and group 3 have the same pattern and are difficult to distinguish, but group 1 has a different pattern. The mean functions for all three groups are different as well, where groups 2 and 3 are closer but group 1 has different mean function. Figure \ref{fig:simulation-3}(a) shows the differences between groups are across all time points, but for Figure
\ref{fig:simulation-3}(b), most of the time points, the three groups are identical, except time 5 to time 15. Compared to group 2 and group 3, group 1 has different patterns but similar means.

\subsection{Simulation Results and Comparison}

\begin{table}[h]
    
    \caption{Multi-Class Classification (left) and variable selection (1st discriminant: middle, 2nd discriminant: right) test metrics for simulated data based on 100 repetitions. The highest metrics are in bold.}
    \label{tab:simresults-multi}
    \centering
    \resizebox{\columnwidth}{!}{
    \begin{tabular}{c|c|ccc|ccc|ccc}
    \hline
    \hline
         &       & \multicolumn{3}{c|}{Classification} & \multicolumn{3}{c|}{Feature Selection 1st} & \multicolumn{3}{c}{Feature Selection 2nd} \\
    Case &  Model & W.Sens & W.Spec & W.F-1 & Sens & Spec & F-1 & Sens & Spec & F-1\\
    \hline
    Different in    &   MFLDA-I   &  \textbf{1.00} &  \textbf{1.00} &  \textbf{1.00} & 0.65 & \textbf{1.00} &  \textbf{0.79} & \textbf{0.75} & \textbf{1.00} &  \textbf{0.86}\\
    all [0,40]   &   PLDA   &  0.99 & 0.99  &  0.99 &  0.73 &  0.30 &  0.22 &  0.73 &  0.30 &  0.24 \\
       &   FPCA+PLDA   & \textbf{1.00}  & \textbf{1.00} &  \textbf{1.00} & \textbf{1.00}  & 0.00 & 0.33 & -  & - & - \\
    \hline
    Different only   &   MFLDA-I   &  \textbf{1.00}  &  \textbf{1.00} &  \textbf{1.00} &  \textbf{1.00}  & 0.00  &  \textbf{0.33} &  \textbf{1.00}  & 0.00  & \textbf{0.33} \\
    in [5,15]   &   PLDA   &  0.82  &  0.91 &  0.82 &  0.95  &  \textbf{0.05} &  0.32 &  0.88  &  \textbf{0.13} &  0.32\\
       &   FPCA+PLDA   &  0.97 & 0.98  & 0.97 & \textbf{1.00}  & 0.00 & \textbf{0.33} & -  & - & - \\
    \hline
    Different in    &   MFLDA-I   & \textbf{1.00}  & \textbf{1.00}  & \textbf{1.00}  & 0.90  &  \textbf{1.00} &  \textbf{0.95} &  \textbf{1.00}  &  0.00 &  \textbf{0.33} \\
    random 10 time   &   PLDA   &  0.99 & 0.99  &  0.99 &  0.56 &  0.48 &  0.25 &  0.54 & \textbf{0.50}  &  0.25\\
     points   &   FPCA+PLDA   & 0.97  & 0.98  &  0.97 & \textbf{1.00}  & 0.00 & 0.33 & -  & - & - \\
    \hline
    Different in random   &   MFLDA-I   & \textbf{1.00}  & \textbf{1.00}  &  \textbf{1.00} & 0.55  & \textbf{1.00}  &  \textbf{0.71} &  \textbf{1.00} &  \textbf{0.95} & \textbf{0.91}\\
    time and random    &   PLDA   & 0.99  &  0.99 & 0.99 & 0.70  & 0.33 & 0.28 &  0.70 & 0.33 & 0.28\\
    length of time  &   FPCA+PLDA   & 0.97  & 0.99  &  0.97 & \textbf{1.00}  & 0.00 & 0.33 & -  & - & - \\
    \hline
    \hline
    \end{tabular}}

\end{table}   

\begin{figure}[h]
    \centering
        \includegraphics[width=\textwidth]{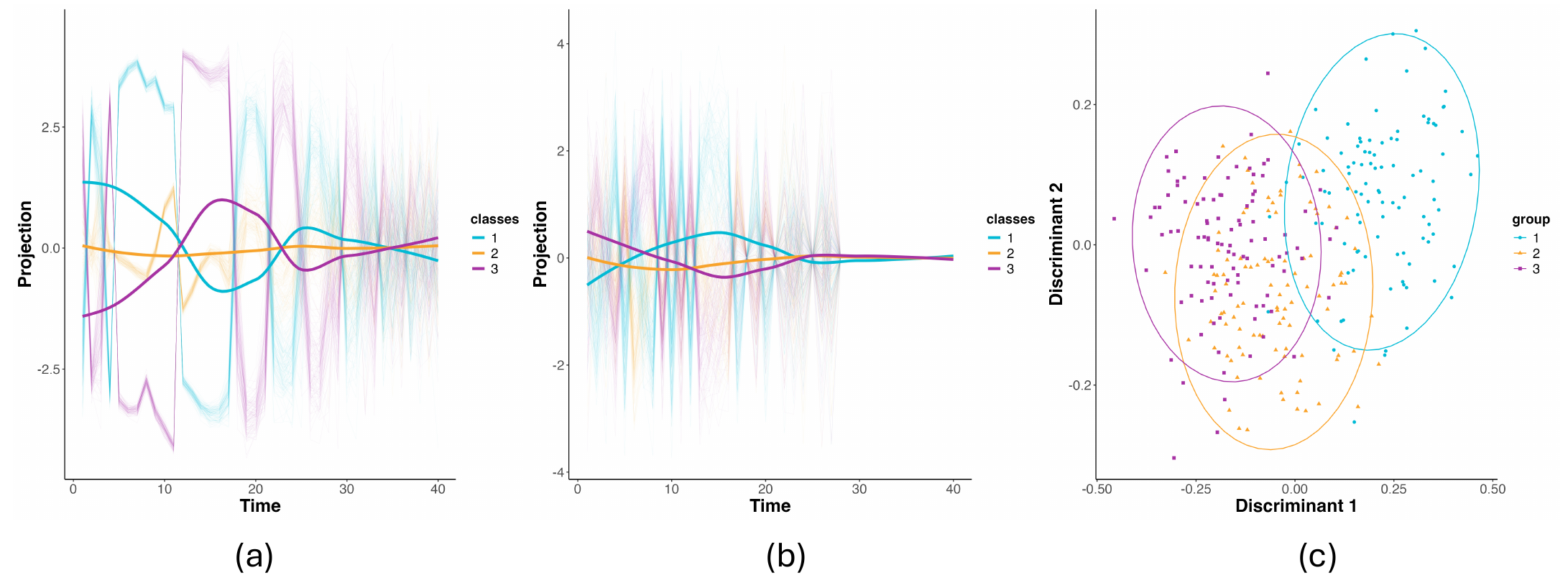}
        \caption{Discriminant Plot of Multi-Class: (a-b) Discriminant plot for first and second discriminant vector. Thin lines represent projection of each subjects and thick lines are LOESS curves colored by classes; (c) Average discriminant score across time colored by groups}
        \label{fig:discriminant-multi}
\end{figure}

We compared the performance of our method by testing its sensitivity, specificity, and F1 score. Table \ref{tab:simresults-multi} displays the results of the classification metrics. For evaluating three-class classification performance, we utilize weighted sensitivity, weighted specificity, and the weighted F1 score. These metrics are designed to address potential challenges arising from imbalanced group distributions, such as overemphasis on majority classes or diminished attention to minority classes. By assigning appropriate weights based on class proportions, these metrics ensure a more balanced evaluation, reducing the risk of misleading performance interpretations and enabling fairer comparisons. Table \ref{tab:simresults-multi} shows that our proposed method MFLDA-I has a generally good performance compared to the other two methods. The FLDA Best method is not displayed here since it is not applicable for three-class functional data. Figure \ref{fig:discriminant-multi} visualizes the discriminant plot. In the discriminant plot, a smoothed LOESS curve is produced. We can find in (a) that the two groups are quite separated all the time, while in (b) the two groups are separated mostly during the time period [5,15], and for most of the remaining time, the two groups are not that distinguishable. The plot shows the consistency with Case 2. Figure \ref{fig:discriminant-multi}(c) shows the average discriminant score across time for the first and second discriminant vector for case 2.

Our proposed MFLDA-I method also generally finds the true variables that distinguish between the three classes. Table \ref{tab:simresults-multi} (middle) compares the first discriminant vector. In most cases, our method consistently identifies the majority of true variables, achieving an F1 score near 0.80. However, in Case 2, where class discrimination relies on variables that differ only between time points 5 and 15, the F1 score drops to 0.33. Similarly, in Case 4, where features vary randomly in both timing and duration, the F1 score is 0.71. Across all cases, PLDA shows moderate sensitivity but suffers from low specificity, resulting in weaker F-1 scores. FPCA + PLDA achieves perfect sensitivity across all cases but sacrifices specificity entirely, limiting its F-1 scores to 0.33. In these cases, our proposed method can utilize the best interpretable variable selection for the first discriminant vector.

Table \ref{tab:simresults-multi}(right) compares the second discriminant vector. Our MFLDA-I model performs well in Case 1 and Case 4, achieving F1 scores of 0.86 and 0.91, respectively. In both cases, MFLDA-I outperforms in terms of sensitivity and specificity. However, in Cases 2 and 3, while MFLDA-I demonstrates better sensitivity, its lower specificity compared to PLDA results in comparable F1 scores. For two-step FPCA + PLDA method, it can only select the first discriminant vector and fail to selected the second discriminant vector, thus thereby showing no results.


\section{More on IBD data}
\subsection{iHMP IBD Dataset} 
Data that motivated the MFLDA method can be downloaded from \href{https://ibdmdb.org/}{The Inflammatory Bowel Disease Multi'omics Database (https://ibdmdb.org/)}. Participants in this IBD initiative take part from the Center for the Study of Inflammatory Bowel Diseases at Massachusetts General Hospital, Emory University Hospital, Cincinnati Children's Hospital, and Cedars-Sinai Medical Center. The IBD study is supported by an interdisciplinary team of experts from these and other institutions, who are responsible for data generation and analysis.

\subsection{Data Processing}
We briefly describe how metagneomics data were obtained. Please refer to \cite{IBDprotocal} for more detials. Metagenomic DNA samples are quantified using the Quant-iT PicoGreen dsDNA Assay (Life Technologies) and subsequently adjusted to a concentration of 50 pg/µL. For library preparation, 100-250 pg of DNA is utilized with the Nextera XT DNA Library Preparation Kit (Illumina), adhering to the manufacturer’s protocol with adjusted reaction volumes. Before sequencing, libraries are pooled by combining equal volumes (200 nl) from batches of 96 samples. The insert sizes and concentrations of each pooled library are then assessed using the Agilent Bioanalyzer DNA 1000 kit (Agilent Technologies). Sequencing is performed on the Illumina HiSeq platform using a 2x101 paired-end format, generating approximately 10 million paired-end reads per library. Sequencing, de-multiplexing and the production of BAM and Fastq files are conducted utilizing the Picard suite \citep{picard}.

\begin{figure}[h]
    \centering
        \includegraphics[width=\textwidth]{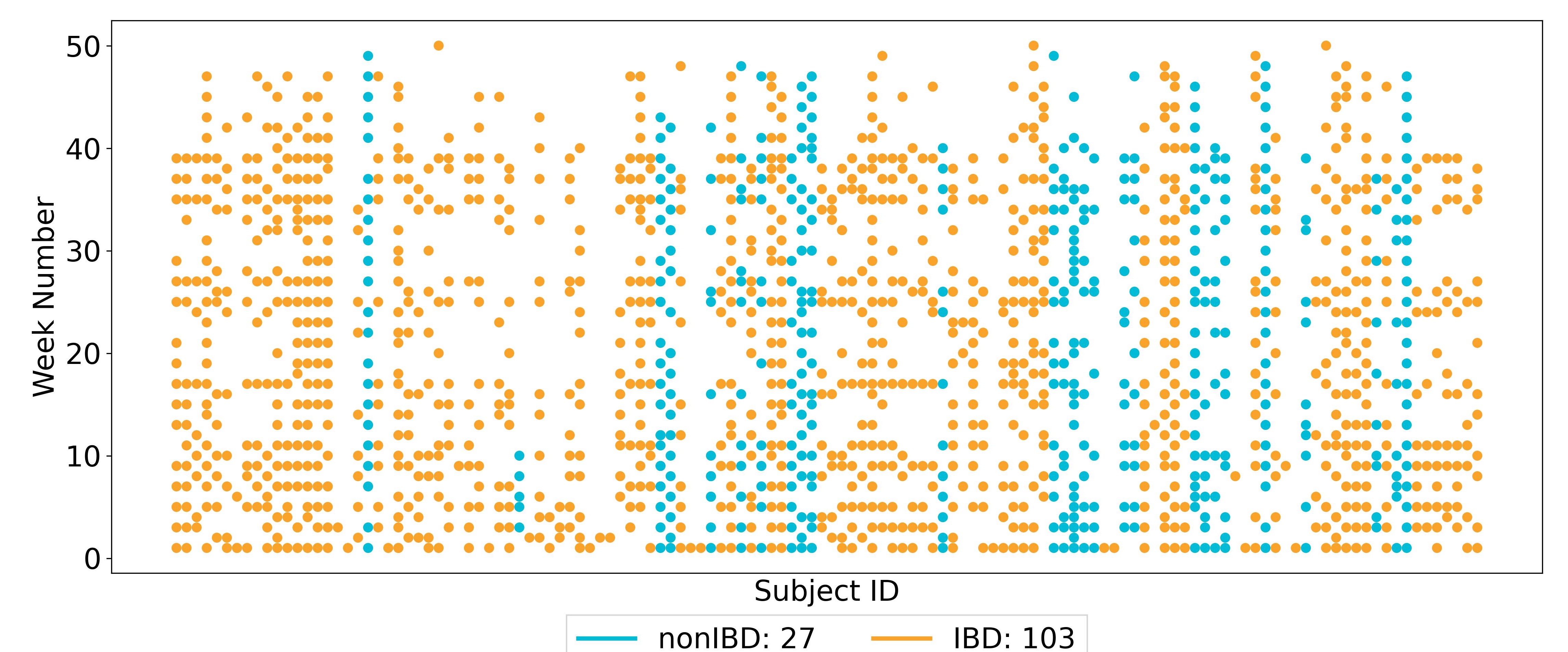}
        \caption{IBD Subject Data Submission Across 50 Weeks. Dots represents whether the subject contributed data at specific weeks. Blank space represents the missing contribution from subjects. Legend shows colors of IBD status and number of subjects in each status.}
        \label{fig:IBDtwoclass}
\end{figure}

The iHMP IBD dataset contains 130 subjects, where subjects submit data in 50 weeks. Figure \ref{fig:IBDtwoclass} shows the data submission for each subject, where it is noticeable that some subjects have missing data, making the functional data sparse. For this analyses, we combine both CD and UC cases as ``IBD" group. Thus, this becomes a binary classification problem.

Table \ref{tab:IBD2demo} displays two classes IBD data baseline statistics including research sites, and subjects' age at diagnosis. Before preprocessing, the metagenomics data contained pathway abundances of 22,113 gene pathways. We note that most of the pathways contain zeros, but our proposed method would be better at classifying groups with abundances that contain variations (i.e. pathways with most zeros are flat-line across times). Preprocessing steps are inspired by \cite{zhao2021tpm, maza2016papyro, jain2024deepida} which consisting of (i) keeping pathways that have less than 80\% zeros; (ii) adding a pseudo count of 1 to each data value (this ensures the logarithm transformation in later steps); (iii) applied central log ratio transformation; (iv) plotting the histogram of variances and filtering out pathways with low variance (less than the first 5\% quantile) across all collected subjects. After the preprocessing steps, the number of pathways remaining for the metagenomics is 1622.

\begin{table}[h] \centering \renewcommand*{\arraystretch}{1.1}\caption{Summary Statistics for Two Classes IBD Data}\resizebox{\textwidth}{!}{
\begin{tabular}{lrrrrrrl}
\hline
\hline
group & \multicolumn{3}{c}{IBD} & \multicolumn{3}{c}{nonIBD} &   \\ 
 Variable & \multicolumn{1}{c}{N} & \multicolumn{1}{c}{Mean} & \multicolumn{1}{c}{SD} & \multicolumn{1}{c}{N} & \multicolumn{1}{c}{Mean} & \multicolumn{1}{c}{SD} & \multicolumn{1}{c}{Test} \\ 
\hline
site\_name & 103 &  &  & 27 &  &  & $\chi^2=12.75^{**}$ \\ 
... Cedars-Sinai & 32 & 31\% &  & 1 & 4\% &  &  \\ 
... Cincinnati & 24 & 23\% &  & 9 & 33\% &  &  \\ 
... Emory & 10 & 10\% &  & 1 & 4\% &  &  \\ 
... MGH & 24 & 23\% &  & 13 & 48\% &  &  \\ 
... MGH Pediatrics & 13 & 13\% &  & 3 & 11\% &  &  \\ 
Age.at.diagnosis & 103 & 21 & 12 & 27 & 21 & 0 & F$=0.019^{}$\\ 
\hline
\hline
\multicolumn{8}{l}{Statistical significance markers: * p$<0.1$; ** p$<0.05$; *** p$<0.01$}\\ 
\end{tabular}
}
\label{tab:IBD2demo}
\end{table}

\subsection{Linear Mixed Model (LMM)}
Linear mixed models (LMMs) extend traditional linear models by incorporating fixed and random effects to account for dependencies in data obtained from repeated measurements \citep{liu2024incorporating}. We apply LMMs to analyze longitudinal data from the IBD study to conduct a comprehensive assessment of the differential abundance of pathways. We performed LMM prior to applying our method in order to use variables that have potential for discriminating  between groups. For this purpose, we compared full and  null LMM models to identify microbial pathways that significantly discriminante between the two groups: (i) a null model and (ii) a full model. 

Null Model: Feature\_value $\sim$ Time + Age at Diagnosis + (1 + time$|$ Subject id) + (1$|$ Site Name)

Full Model: Feature\_value $\sim$ IBD\_Group + Time + Age at Diagnosis + (1 + time$|$ Subject id) + (1$|$ Site Name)

In the null model, the longitudinal variable is associated with a fixed variable (e.g., time and age at diagnosis), random intercept, random slope and random effect of research sites. The full model extends the null model by incorporating the IBD groups of the sample as an additional variable. Subsequently, an ANOVA comparison is performed between the full and null models to identify statistically significant variables (p-value $< 0.05$) that effectively discriminate between the considered groups.

\begin{figure}[h]
    \centering
        \includegraphics[width=\textwidth]{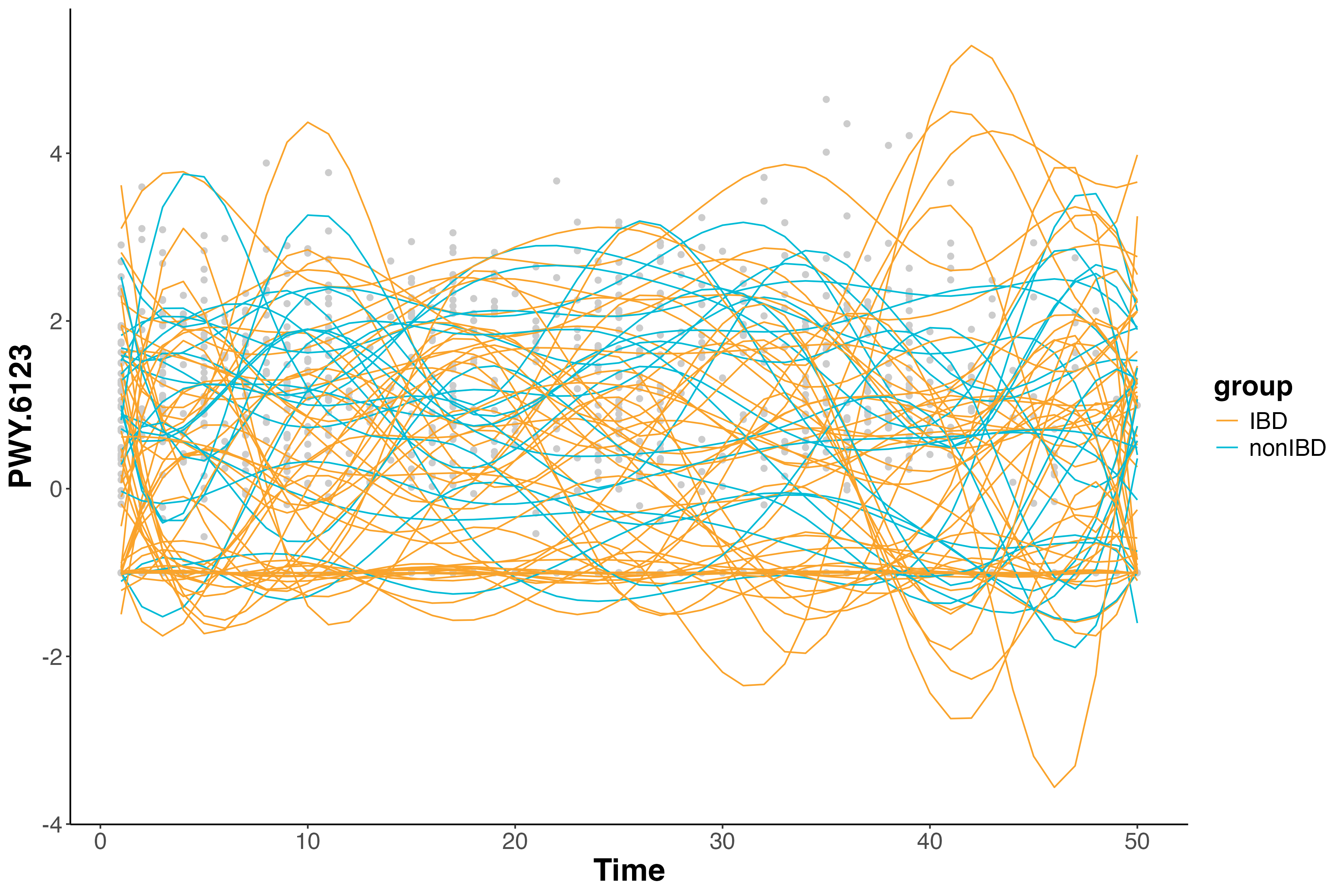}
        \caption{IBD two class example of one microbial pathway. Standardized pathway abundance changes across 50 weeks. Dots represents the original abundance and lines represents b-spline estimation of each subject colored by IBD status.}
        \label{fig:twoexample}
\end{figure}

By applying LMMs, statistically significant longitudinal variables are considered for the next step. Figure \ref{fig:twoexample} shows an example of a significant variable. The x-axis represents weeks and the y-axis represents standardized pathway abundances after preprocessing. The gray dots represent the observed data points. As the example shows, observed data are quite sparse with multiple missing. The colored lines represent the estimation of the b-spline by the IBD groups. To establish the b-spline estimation, we use a minimum number of 8 time points for each subject. Meanwhile, 30 subjects do not have enough information about time points for the b-spline estimation. Therefore, a total of 100 subjects and the top 200 features are included in the analysis.

\subsection{Binary Outcome Results (IBD vs. Non-IBD)}

We begin by performing a binary classification between BD and Non-IBD assuming pathways are time independent, and comparing our MFLDA-I method with several established approaches, including FLDA for multiple functional data by Gardner-Lubbe \citep{gardner2021linear}, PLDA \citep{witten2011penalized}, and FPCA \citep{yao2005functional}, using 5-fold cross-validation. The sensitivity, specificity, F-a score, accuracy, balanced accuracy and MCC score for different methods are presented in Table \ref{tab:IBD2results}.

\begin{table}[h]
\caption{Classification Metrics for Predicting 2 IBD Groups based on 5-Fold Cross-Validation. The highest metrics are in bold.}
    \label{tab:IBD2results}
    \centering
    \begin{tabular}{r|rrrr}
    \hline
    \hline
    Metrics & MFLDA-I & PLDA & FLDA& FPCA+PLDA\\
    \hline
    Sensitivity   &   0.784 & 0.975 & \textbf{1.000} & 0.994\\
    Specificity  &   \textbf{0.615} & 0.002 & 0.000 & 0.004\\
    F-1 Score   &   0.817 & \textbf{0.845} & 0.788 & 0.839\\
    Accuracy   &   \textbf{0.740} & 0.734 & 0.650 & 0.723\\
    Balanced Accuracy   &  \textbf{0.700} & 0.498 & 0.500 & 0.499\\
    MCC  &  \textbf{0.375} & -0.013 & 0.000 & -0.012\\
    \hline
    \hline
    \end{tabular}
\end{table}

Compared with other methods, our proposed MFLDA-I works well, especially in Specificity and MCC. A high sensitive score indicates a lower false negative rate, reducing the probability of missed disease cases. Specificity refers to a test's ability to correctly identify individuals who do not have the disease, designating them as negative, and a high specific score results in fewer false positive outcomes. Based on Table \ref{tab:IBD2results}, one can notice that MFLDA-I is better at identifying the true positive subjects. The MCC score is in the range of -1 to 1, and a higher MCC score is achieved only if the classifier demonstrates high values in all four fundamental rates of the confusion matrix: sensitivity, specificity, precision, and negative predictive value. Meanwhile, in this binary problem, the number of ``IBD'' subjects (74 subjects) is three times that of the number of ``non-IBD'' cases (26 subjects), and it is more reliable to compare the balanced accuracy than the accuracy. MFLDA-I provides a balanced accuracy of 0.7, whereas other methods are less than 0.5 (i.e., nearly a random guess). Our method selected 57 significant features, demonstrating its effectiveness in feature selection, whereas FLDA lacks interpretability. PLDA, which evaluates the features at each of the 50 time points, selected all the features at only two time points, while the rest chose none; however, even when selecting all features, its classification performance was not improved. The two-step FPCA+PLDA approach also failed to select any features.

\subsubsection{Microbiome Pathways Selected by MFLDA-I}

\begin{figure}[h]
    \centering
        \includegraphics[width=\textwidth]{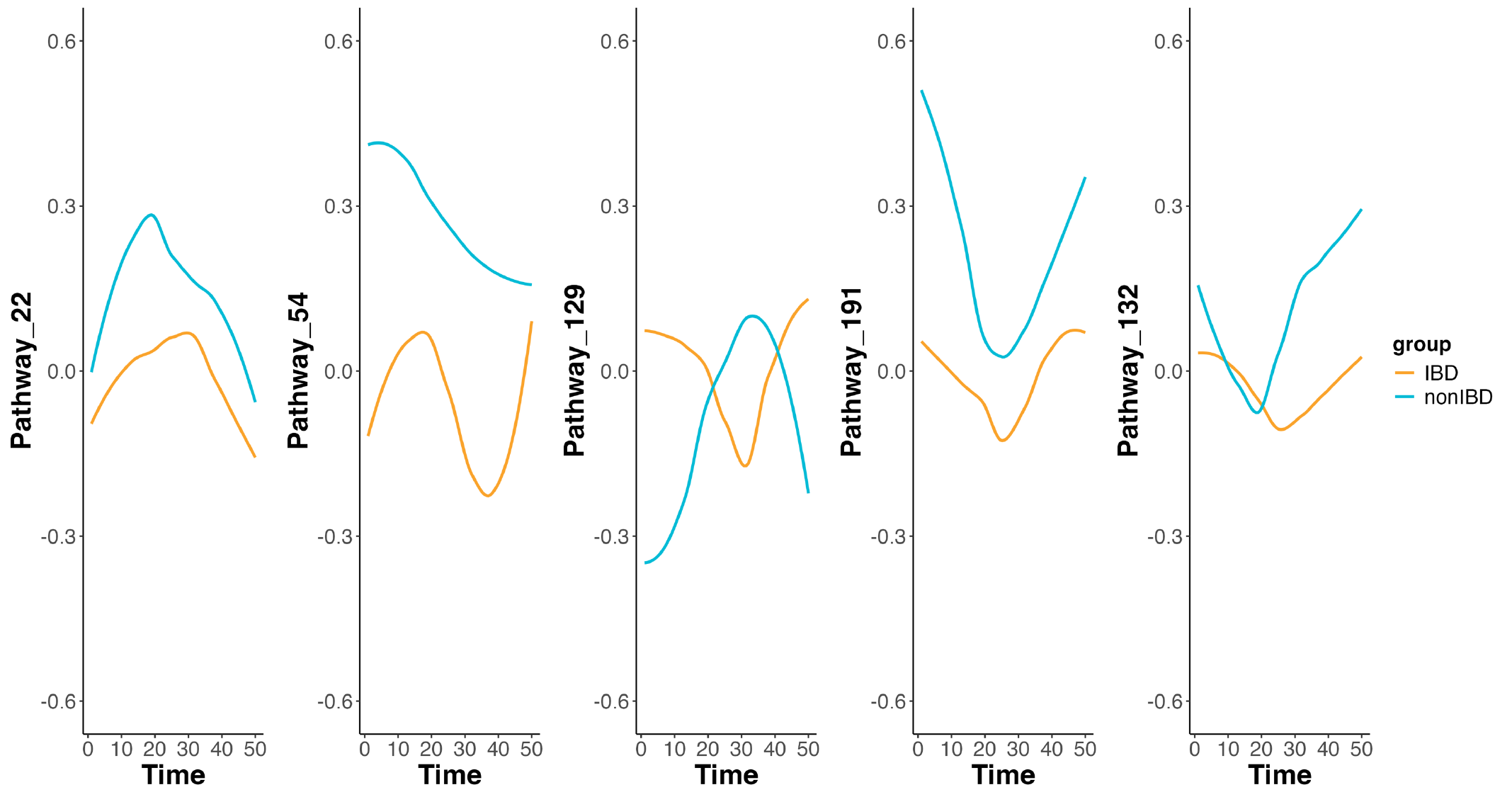}
        \caption{5 Selected Pathways for IBD vs. Non-IBD Status. Lines represented average time series curves of standarized pathway abundance for each IBD status group. MFLDA-I is capable to identify pathways that distinguish between IBD and non-IBD group.}
        \label{fig:binary_pathway}
\end{figure}

\begin{figure}
\centering
\animategraphics[loop, controls, width=15cm]{2}{animation_ibd2/pathway}{1}{57}
\caption{Selected Features for IBD vs. Non-IBD Status. Lines represented average time series curves of standardized feature abundance for each IBD status group. Click to start the figure animation and use controls to navigate, start, and stop.}
\label{fig:pathways_vis_2}
\end{figure}

Analysis of metagenomic pathways (features) distinguishing between patients with and without IBD reveals several key mechanisms influencing gut health and disease progression. In our MFLDA methods, we have selected 57 features. Figure \ref{fig:binary_pathway} shows five of the selected microbial pathways discriminating by IBD status over time. The five pathway names are:
\begin{itemize}
    \item DTDPRHAMSYN.PWY..dTDP.L. rhamnose.biosynthesis.I.g \_Bacteroides.s \_Bacteroides \_vulgatus \_CAG \_6
    \item PPGPPMET.PWY..ppGpp.biosynthesis.unclassified
    \item PWY.6731..starch.degradation.III
    \item SULFATE.CYS.PWY..superpathway.of.sulfate.assimilation.and.cysteine.biosynthesis .unclassified
    \item PWY.6803..phosphatidylcholine.acyl.editing.unclassified
\end{itemize}
The two colors represent IBD or non-IBD status for the average features abundance of all subjects at all times. From these curves, our MFLDA model can efficiently identify features that either have different means [e.g., Pathway\_22, Pathway\_191], different trends [e.g., Pathway\_129], or have both different means and different trends [e.g., Pathway\_54]. Our MFLDA also identified features that cannot distinguish between two groups at the beginning but differ at the rest times [e.g., Pathway\_132]. Figure \ref{fig:pathways_vis_2} shows all selected pathways.

The features selected involve specific gut bacteria, such as \textit{Blautia} \citep{liu2021blautia}, \textit{Ruminococcus} \citep{henke2019ruminococcus}, \textit{Alistipes} \citep{parker2020genus}, and \textit{Bacteroides} \citep{zhou2016lower}, highlight the role of microbial composition and dysbiosis in IBD. Other relevant gut microbiota processes involve amino acid metabolism \citep{wu2022role, li2014glutamate} and short-chain fatty acid (SCFA) production \citep{basson2016mucosal} are key for protein synthesis, immune regulation, mucosal health, epithelial integrity, and providing energy to colonocytes. These microbiome pathways significantly influence gut barrier function, immune responses, and inflammatory regulation, distinguishing IBD patients from healthy individuals.

\subsection{More on Multi-Class Problem (CD vs. UC vs. Non-IBD)}
Crohn's disease (CD) and ulcerative colitis (UC) are both forms of inflammatory bowel disease (IBD) characterized by chronic inflammation of the gastrointestinal tract (GI), sharing symptoms such as abdominal pain, diarrhea, weight loss, fatigue, and extraintestinal manifestations such as joint pain and skin rashes \citep{abraham2015symptom}. Both conditions have genetic predispositions and involve abnormal immune responses to gut microbiota, with similar treatment approaches including anti-inflammatory drugs and biologics \citep{basso2014association}. However, CD can affect any part of the GI tract, typically the ileum and colon, with patchy, transmural inflammation leading to complications like strictures and fistulas, while UC is confined to the colon and rectum with continuous mucosal inflammation, potentially causing toxic megacolon and increased colon cancer risk \citep{ardizzone2000altered, yantiss2006diagnostic, gajendran2019comprehensive}. 

\begin{table}[h] \centering \renewcommand*{\arraystretch}{1.1}\caption{Summary Statistics for Three Classes IBD Data}\resizebox{\textwidth}{!}{
\begin{tabular}{lrrrrrrrrrl}
\hline
\hline
group & \multicolumn{3}{c}{CD} & \multicolumn{3}{c}{nonIBD} & \multicolumn{3}{c}{UC} &   \\ 
 Variable & \multicolumn{1}{c}{N} & \multicolumn{1}{c}{Mean} & \multicolumn{1}{c}{SD} & \multicolumn{1}{c}{N} & \multicolumn{1}{c}{Mean} & \multicolumn{1}{c}{SD} & \multicolumn{1}{c}{N} & \multicolumn{1}{c}{Mean} & \multicolumn{1}{c}{SD} & \multicolumn{1}{c}{Test} \\ 
\hline
site\_name & 65 &  &  & 27 &  &  & 38 &  &  & $\chi^2=14.244^{*}$ \\ 
... Cedars-Sinai & 20 & 31\% &  & 1 & 4\% &  & 12 & 32\% &  &  \\ 
... Cincinnati & 17 & 26\% &  & 9 & 33\% &  & 7 & 18\% &  &  \\ 
... Emory & 7 & 11\% &  & 1 & 4\% &  & 3 & 8\% &  &  \\ 
... MGH & 13 & 20\% &  & 13 & 48\% &  & 11 & 29\% &  &  \\ 
... MGH Pediatrics & 8 & 12\% &  & 3 & 11\% &  & 5 & 13\% &  &  \\ 
Age.at.diagnosis & 65 & 20 & 10 & 27 & 21 & 0 & 38 & 24 & 15 & F$=1.938^{}$\\ 
\hline
\hline
\multicolumn{11}{l}{Statistical significance markers: * p$<0.1$; ** p$<0.05$; *** p$<0.01$}\\ 
\end{tabular}
}
\label{tab:IBDthreeclass-demo}
\end{table}

\begin{figure}[h!]
    \centering
        \includegraphics[width=\textwidth]{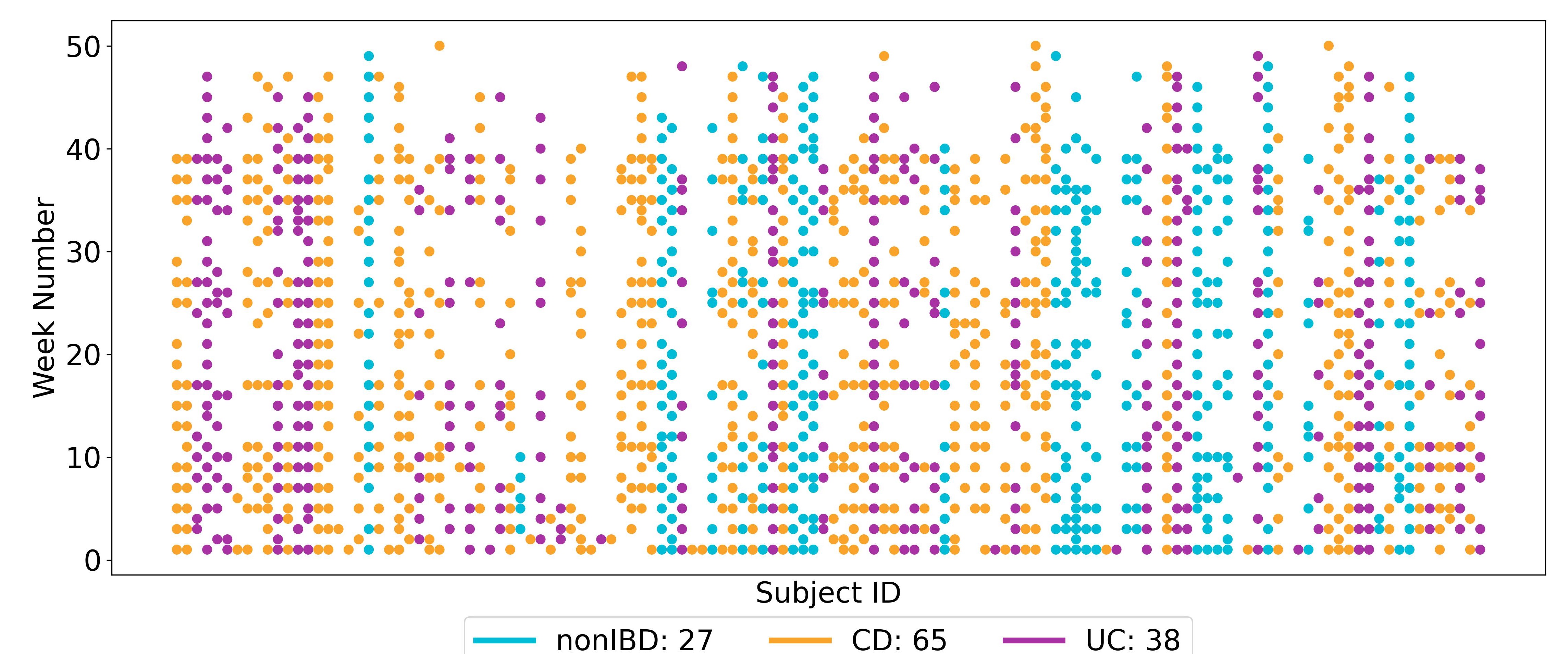}
        \caption{IBD Subject Data Submission Across 50 Weeks. Dots represents whether the subject contributed data at specific weeks. Blank space represents the missing contribution from subjects. Legend shows colors of IBD status and number of subjects in each status.}
        \label{fig:IBDthreeclass}
\end{figure}

Figure \ref{fig:IBDthreeclass} provides a similar dots plot showing the sparsity of the data. The empty dots indicate no data entry for one visit of one subject. As we can see here, there are many missing data. We exclude 30 individuals with less than 8 measurements and kept the top 200 features for further analysis.

\begin{figure}[h]
    \centering
        \includegraphics[width=\textwidth]{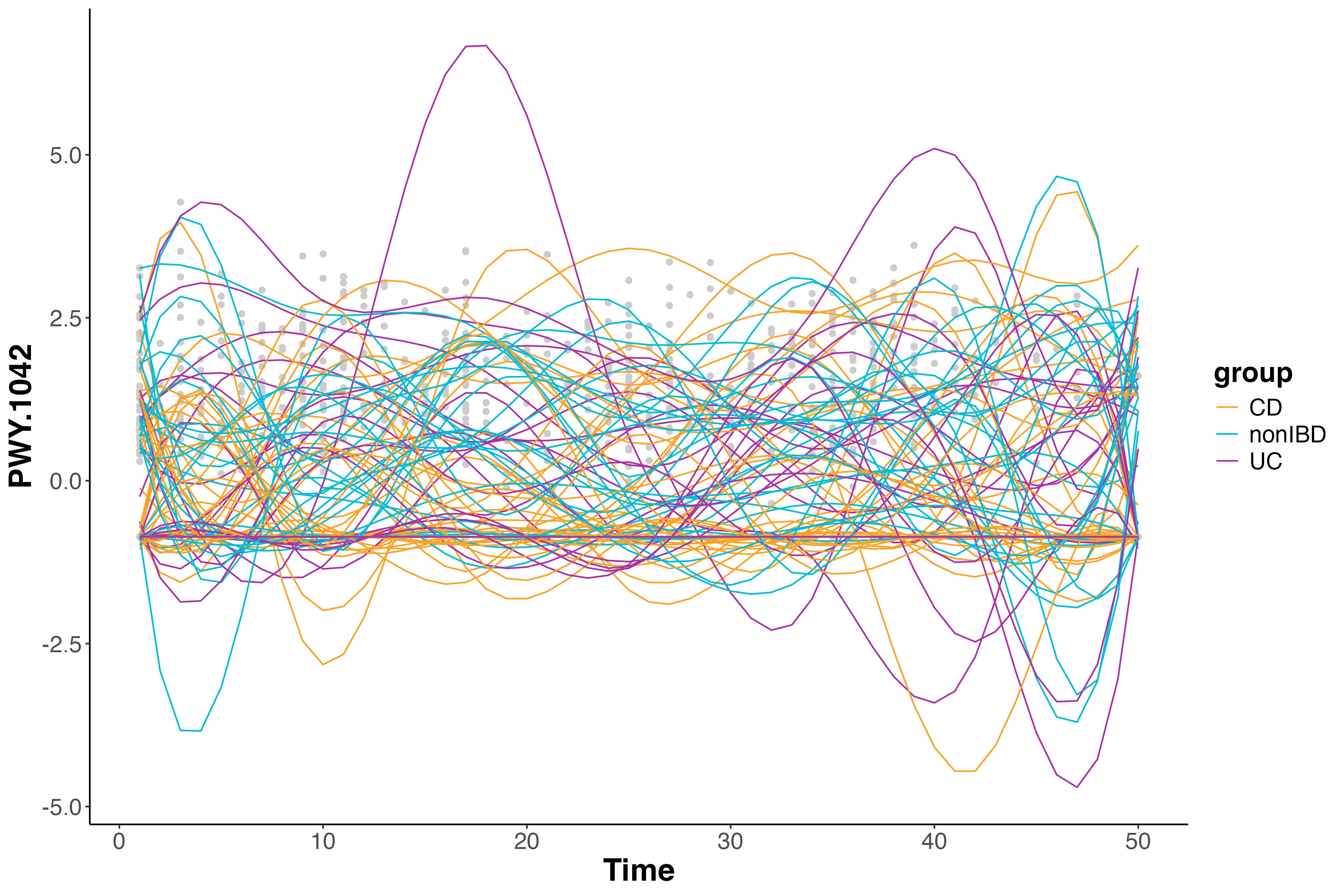}
        \caption{IBD three class example of one pathway. Dots represents the original abundance and lines represents b-spline estimation of each subject colored by IBD status.}
        \label{fig:threeexample}
\end{figure}

In Figure \ref{fig:threeexample}, we show sample curves for one microbial pathway. The gray dots represent the true observation and the colored curves represent the smoothed function, by IBD group,  for all samples. From this figure, it appears that the groups are difficult to distinguish based on this pathway. Therefore, it is not surprising that the classification performance of our method and the other methods are not high compared to the binary classification problem (Table 3, main text). 

\subsubsection{Microbiome Pathways Selected by MFLDA-I}
MFLDA-I selects 59 microbiome pathways that can distinguish between subjects with UC, CD, and non-IBD. Figure \ref{fig:pathways_vis_3} is an animation of all selected microbial pathways, with colored lines showing the average feature abundance by group. In some pathway profiles, CD and UC exhibit similar patterns [e.g., Pathway\_45, Pathway\_50] and are separated from non-IBD, which may have different means or opposite patterns. In some cases, non-IBD is distinguishable from UC but not CD [e.g., Pathway\_142]. MFLDA effectively identifies pathways that distinguish these groups.

\begin{figure}
\centering
\animategraphics[loop, controls, width=15cm]{2}{animation_ibd3/pathway}{1}{59}
\caption{Selected Features for CD vs. UC vs. Non-IBD Status. Lines represented average time series curves of standardized feature abundance for each IBD status group. Highlighted pathways are mentioned in main text. Click to start the figure animation and use controls to navigate, start, and stop.}
\label{fig:pathways_vis_3}
\end{figure}

Certain microbiome processes related to the human gut microbiota, such as mucin degradation, amino acid metabolism, and peptidoglycan synthesis, can help distinguish IBD subtypes (UC and CD). Using our method, we identified relevant microbiome pathways. The \textit{Akkermansia muciniphila} pathway, linked to mucin degradation, is disrupted in UC, reflecting differences in gut lining integrity and inflammation levels \citep{zheng2023role}. Several pathways related to peptidoglycan recognition are important in immune modulation and may contribute to inflammatory responses in CD and/or UC \citep{zulfiqar2013genetic, inohara2003host}. These pathways help differentiate microbial activity and inflammation in UC and CD from each other and from non-IBD conditions. Other selected pathways could be further studied for their potential roles in IBD.

\subsection{Comparison Between Microbial Pathways Selected in the Binary and Multi-Class Classification Problems using MFLDA-I}
Both microbial pathways for distinguishing between IBD and Non-IBD or among UC, CD, and Non-IBD exhibit significant overlaps and distinctions in microbial metabolic pathways. They share common pathways such as those for dTDP-L-rhamnose biosynthesis \citep{li2022capsular}, L-histidine biosynthesis and degradation \citep{wu2022role}, pantothenate \citep{schalich2024human} and coenzyme A biosynthesis \citep{ellestad1976pantothenic}, gluconeogenesis \citep{eriksson1983splanchnic}, peptidoglycan biosynthesis and maturation \citep{zulfiqar2013genetic, inohara2003host}, and L-isoleucine biosynthesis \citep{xie2021systematic}. However, they diverge in their specificity and diversity. Variables for distinguishing IBD and non-IBD status covers a broader array of pathways, including unique ones like L-citrulline biosynthesis \citep{middleton1993increased} and the methylerythritol phosphate pathway \citep{schirmer2019microbial}, and features a more diverse set of microbial species such as \textit{Parasutterella} \citep{chen2018parasutterella} and \textit{Dialister} \citep{joossens2011dysbiosis}. In contrast, variables for distinguishing CD, UC, and non-IBD status emphasize pathways specific to \textit{Alistipes putredinis} \citep{nomura2021bacteroidetes, wu2024changes} and \textit{Blautia torques} \citep{liu2021blautia} and include unique pathways performing processes such as incomplete reductive TCA cycle \citep{ooi2011gc}, sulfur metabolism \citep{teigen2019dietary, metwaly2020integrated}, and energy metabolism \citep{ozsoy2022role} in invertebrates. Furthermore, variables to distinguish CD, UC, and non-IBD include specific pathways for vitamin and coenzyme biosynthesis, such as thiamin diphosphate \citep{frank2015thiamin} and flavin, and offer variants of common pathways like glycolysis in plant cytosol and various forms of thiamin diphosphate biosynthesis, highlighting a broader spectrum of metabolic adaptations.

\begin{table}[h]
\caption{Microbiome Pathways (shortened names due to space constraint) Selected by Each Method: common pathways selected are in red, vitamin related pathways are highlighted.\\
    * PLDA selected all 200 pathways in 2 out of 50 time points (only show partial)\\
    $\dagger$ PLDA selected all 200 pathways in 17 out of 50 time points (only show partial)\\
    $\mathsection$ FPCA+PLDA selected all 200 pathways (only show partial)}
    \label{tab:var-compare}
    \centering
    \begin{tabular}{p{1.3cm}|p{4.5cm}|p{4.1cm}|p{4.1cm}}
    Method & MFLDA-I & PLDA & FPCA+PLDA\\
    \hline
    \hline
    IBD vs. non-IBD   &  \fontsize{4}{4}\selectfont\label{var-binary-SFLDA}
UNINTEGRATED- Blautia- Ruminococcus, Branched- Chain- AA- Syn- Alistipes, Pyruvate- Butanoate- Fermentation, Citrulline- Biosynthesis, dTDP- Rhamnose- Alistipes, dTDP- Rhamnose- Bacteroides- Fragilis, dTDP- Rhamnose- Bacteroides- Vulgatus, dTDP- Rhamnose- Dorea, Fucose- Degradation, Histidine- Degradation- Bacteroides- Xylanisolvens, Histidine- Biosynthesis- Bacteroides- Fragilis, Isoleucine- Threonine- Biosynthesis, Methylerythritol- Pathway, \hl{Acetylneuraminate- Degradation}, \highlight{yellow}{red}{\textbf{CoA- Biosynthesis- Unclassified}}, ppGpp- Biosynthesis, C4- Carbon- Assimilation- NADP- ME, Histidine- Degradation- III, Isoleucine- Biosynthesis- Blautia, Isoleucine- Biosynthesis- Unclassified, Glutaryl- CoA- Degradation- Faecalibacterium, \textcolor{red}{\textbf{Glutamate- Glutamine}}, Acetyl- CoA- Butanoate- II, \hl{Menaquinol- 9- Biosynthesis}, Palmitate- Biosynthesis, Cis- Vaccenate- Bacteroides- Uniformis, Stearate- Biosynthesis, 5- Aminoimidazole- Biosynthesis- Parasutterella, Inosine- Biosynthesis- Alistipes, Inosine- Biosynthesis- Dialister, 5- Aminoimidazole- Superpathway- Parasutterella, Adenine- Salvage- Bacteroides- Stercoris, Adenine- Salvage- Eubacterium- Rectale, \textcolor{red}{\textbf{PreQ0- Biosynthesis}}, Starch- Degradation- III, Phosphatidylcholine- Acyl- Editing, Glucose- Xylose- Degradation, C4- Carbon- Assimilation- PEPCK, Pyrimidine- Salvage- Bacteroides- Fragilis, Taxadiene- Biosynthesis- Engineered, Arginine- Biosynthesis- Alistipes, Dihydropterin- Biosynthesis- Bacteroides- Fragilis, \textcolor{red}{\textbf{Gondoate- Anaerobic}}, Oleate- Biosynthesis- Alistipes, \textcolor{red}{\textbf{Peptidoglycan- Maturation- Eubacterium}}, \hl{Pyridoxal- Biosynthesis- Bacteroides- Uniformis}, Phosphatidylglycerol- Biosynthesis- I, Phosphatidylglycerol- Biosynthesis- II, Fermentation- Chlamydomonas, \textcolor{red}{\textbf{Gluconeogenesis- III}}, Mycolate- Biosynthesis- Alistipes, \hl{NAD- Biosynthesis- Bacteroides- Fragilis}, \textcolor{red}{\textbf{Serine- Glycine- Biosynthesis- Alistipes}}, Sulfate- Cysteine- Biosynthesis, \highlight{yellow}{red}{\textbf{Thiamin- Diphosphate- Bacteroides}}, Tryptophan- Biosynthesis- Eubacterium, \textcolor{red}{\textbf{Tryptophan}}   &   \fontsize{4}{4}\selectfont\label{var-binary}
UNINTEGRATED- Blautia- Ruminococcus, Branched- Chain- AA- Syn- Alistipes, Pyruvate- Butanoate- Fermentation, Citrulline- Biosynthesis, dTDP- Rhamnose- Alistipes, dTDP- Rhamnose- Bacteroides- Fragilis, dTDP- Rhamnose- Bacteroides- Vulgatus, dTDP- Rhamnose- Dorea, Fucose- Degradation, Histidine- Degradation- Bacteroides- Xylanisolvens, Histidine- Biosynthesis- Bacteroides- Fragilis, Isoleucine- Threonine- Biosynthesis, Methylerythritol- Pathway, Acetylneuraminate- Degradation, CoA- Biosynthesis- Unclassified, ppGpp- Biosynthesis, C4- Carbon- Assimilation- NADP- ME, Histidine- Degradation- III, Isoleucine- Biosynthesis- Blautia, Isoleucine- Biosynthesis- Unclassified, Glutaryl- CoA- Degradation- Faecalibacterium, Glutamate- Glutamine- Biosynthesis, Acetyl- CoA- Butanoate- II, Menaquinol- 9- Biosynthesis, Palmitate- Biosynthesis, Cis- Vaccenate- Bacteroides- Uniformis, Stearate- Biosynthesis, 5- Aminoimidazole- Biosynthesis- Parasutterella, Inosine- Biosynthesis- Alistipes, Inosine- Biosynthesis- Dialister, 5- Aminoimidazole- Superpathway- Parasutterella, Adenine- Salvage- Bacteroides- Stercoris, Adenine- Salvage- Eubacterium- Rectale, PreQ0- Biosynthesis, Starch- Degradation- III, Phosphatidylcholine- Acyl- Editing, Glucose- Xylose- Degradation, C4- Carbon- Assimilation- PEPCK, Pyrimidine- Salvage- Bacteroides- Fragilis, Taxadiene- Biosynthesis- Engineered, Arginine- Biosynthesis- Alistipes, Dihydropterin- Biosynthesis- Bacteroides- Fragilis, Gondoate- Biosynthesis- Anaerobic, Oleate- Biosynthesis- Alistipes, Peptidoglycan- Maturation- Eubacterium, Pyridoxal- Biosynthesis- Bacteroides- Uniformis, Phosphatidylglycerol- Biosynthesis- I, Phosphatidylglycerol- Biosynthesis- II, Fermentation- Chlamydomonas, Gluconeogenesis- III, Mycolate- Biosynthesis- Alistipes, NAD- Biosynthesis- Bacteroides- Fragilis, Serine- Glycine- Biosynthesis- Alistipes, Sulfate- Cysteine- Biosynthesis, Thiamin- Diphosphate- Biosynthesis, Tryptophan- Biosynthesis- Eubacterium, Tryptophan- Biosynthesis- Unclassified.

\textbf{(*)} & 0 Pathways Selected \\
    \hline
    CD vs. UC vs. non-IBD  & \fontsize{4}{4}\selectfont\label{var-multi-SFLDA}

UNINTEGRATED- Akkermansia- Muciniphila, UNINTEGRATED- Alistipes- Putredinis, UNINTEGRATED- Alistipes- Putredinis- CAG67, \hl{N10- Formyl- Tetrahydrofolate- Alistipes- CAG67}, \hl{N10- Formyl- Tetrahydrofolate- Unclassified, Glycolysis- III- Glucose}, \hl{CoA- Biosynthesis- Ruminococcus}, \highlight{yellow}{red}{\textbf{CoA- Biosynthesis- Unclassified}}, dTDP- Rhamnose- Alistipes- CAG67, Histidine- Biosynthesis, Formaldehyde- Assimilation, Incomplete- Reductive- TCA, \hl{Phosphopantothenate}, \hl{Pantothenate- CoA}, Glycolysis- IV- Plant, CMP- 3- Deoxy- D- Manno, Histidine- Degradation- Alistipes, Pyruvate- Fermentation- Acetate- Blautia, Isoleucine- Biosynthesis- Alistipes, Isoleucine- Biosynthesis- Blautia, Unsaturated- Fatty- Acid- Oxidation, Sulfur- Oxidation- Acidianus, Sulfur- Oxidation- Blautia, \textcolor{red}{\textbf{Glutamate- Glutamine}}, GDP- Mannose, Cis- Vaccenate, Dihydropterin- Alistipes- CAG67, Sucrose- Degradation, \hl{Menaquinol- 8}, Peptidoglycan- Mycobacteria- Alistipes, Peptidoglycan- Enterococcus, \textcolor{red}{\textbf{PreQ0- Biosynthesis}}, \hl{Thiazole- Bacteroides}, \hl{Thiazole- Unclassified}, \hl{Thiamin- Salvage}, TCA- Oxoglutarate, Pyruvate- Fermentation- Isobutanol, Inosine- Biosynthesis- III, Inosine- Biosynthesis, Anaerobic- Metabolism- Invertebrates, Anaerobic- Metabolism, Mannan- Degradation, \hl{Dihydropterin- Chlamydia- CAG67}, \textcolor{red}{\textbf{Gondoate- Anaerobic}}, ADP- Glycero- Manno, Purine- Degradation- Eubacterium, \textcolor{red}{\textbf{Peptidoglycan- Maturation- Eubacterium}}, \textcolor{red}{\textbf{Gluconeogenesis- III}}, Glycolysis- VI- Metazoan, \hl{Flavin- Biosynthesis- Bacteria}, \textcolor{red}{\textbf{Serine- Glycine- Biosynthesis- Alistipes}}, \highlight{yellow}{red}{\textbf{Thiamin- Diphosphate- Bacteroides}}, \hl{Thiamin- Diphosphate}, \hl{Thiamin- Diphosphate- Eukaryotes- Bacteroides}, \hl{Thiamin- Diphosphate- Eukaryotes- Blautia}, \textcolor{red}{\textbf{Tryptophan}}, UDP- Acetyl- Glucosamine- Ruminococcus, Valine- Ruminococcus, Valine- Unclassified

   & \fontsize{4}{4}\selectfont\label{var-multi}

UNINTEGRATED- Akkermansia- Muciniphila, UNINTEGRATED- Alistipes- Putredinis, UNINTEGRATED- Alistipes- Putredinis- CAG67, N10- Formyl- Tetrahydrofolate- Alistipes- CAG67, N10- Formyl- Tetrahydrofolate- Unclassified, Glycolysis- III- Glucose, CoA- Biosynthesis- Ruminococcus, CoA- Biosynthesis- Unclassified, dTDP- Rhamnose- Alistipes- CAG67, Histidine- Biosynthesis, Formaldehyde- Assimilation, Incomplete- Reductive- TCA, Phosphopantothenate, Pantothenate- CoA, Glycolysis- IV- Plant, CMP- 3- Deoxy- D- Manno, Histidine- Degradation- Alistipes, Pyruvate- Fermentation- Acetate- Blautia, Isoleucine- Biosynthesis- Alistipes, Isoleucine- Biosynthesis- Blautia, Unsaturated- Fatty- Acid- Oxidation, Sulfur- Oxidation- Acidianus, Sulfur- Oxidation- Blautia, Glutamate- Glutamine, GDP- Mannose, Cis- Vaccenate, Dihydropterin- Alistipes- CAG67, Sucrose- Degradation, Menaquinol- 8, Peptidoglycan- Mycobacteria- Alistipes, Peptidoglycan- Enterococcus, PreQ0, Thiazole- Bacteroides, Thiazole- Unclassified, Thiamin- Salvage, TCA- Oxoglutarate, Pyruvate- Fermentation- Isobutanol, Inosine- Biosynthesis- III, Inosine- Biosynthesis, Anaerobic- Metabolism- Invertebrates, Anaerobic- Metabolism, Mannan- Degradation, Dihydropterin- Chlamydia- CAG67, Gondoate- Anaerobic, ADP- Glycero- Manno, Purine- Degradation- Eubacterium, Peptidoglycan- Maturation- Eubacterium, Gluconeogenesis- III, Glycolysis- VI- Metazoan, Flavin- Biosynthesis- Bacteria, Serine- Glycine- Alistipes- CAG67, Thiamin- Diphosphate- Bacteroides, Thiamin- Diphosphate, Thiamin- Diphosphate- Eukaryotes- Bacteroides, Thiamin- Diphosphate- Eukaryotes- Blautia, Tryptophan, UDP- Acetyl- Glucosamine- Ruminococcus, Valine- Ruminococcus, Valine- Unclassified.($\dagger$) &  \fontsize{4}{4}\selectfont\label{var-multi}

UNINTEGRATED- Akkermansia- Muciniphila, UNINTEGRATED- Alistipes- Putredinis, UNINTEGRATED- Alistipes- Putredinis- CAG67, N10- Formyl- Tetrahydrofolate- Alistipes- CAG67, N10- Formyl- Tetrahydrofolate- Unclassified, Glycolysis- III- Glucose, CoA- Biosynthesis- Ruminococcus, CoA- Biosynthesis- Unclassified, dTDP- Rhamnose- Alistipes- CAG67, Histidine- Biosynthesis, Formaldehyde- Assimilation, Incomplete- Reductive- TCA, Phosphopantothenate, Pantothenate- CoA, Glycolysis- IV- Plant, CMP- 3- Deoxy- D- Manno, Histidine- Degradation- Alistipes, Pyruvate- Fermentation- Acetate- Blautia, Isoleucine- Biosynthesis- Alistipes, Isoleucine- Biosynthesis- Blautia, Unsaturated- Fatty- Acid- Oxidation, Sulfur- Oxidation- Acidianus, Sulfur- Oxidation- Blautia, Glutamate- Glutamine, GDP- Mannose, Cis- Vaccenate, Dihydropterin- Alistipes- CAG67, Sucrose- Degradation, Menaquinol- 8, Peptidoglycan- Mycobacteria- Alistipes, Peptidoglycan- Enterococcus, PreQ0, Thiazole- Bacteroides, Thiazole- Unclassified, Thiamin- Salvage, TCA- Oxoglutarate, Pyruvate- Fermentation- Isobutanol, Inosine- Biosynthesis- III, Inosine- Biosynthesis, Anaerobic- Metabolism- Invertebrates, Anaerobic- Metabolism, Mannan- Degradation, Dihydropterin- Chlamydia- CAG67, Gondoate- Anaerobic, ADP- Glycero- Manno, Purine- Degradation- Eubacterium, Peptidoglycan- Maturation- Eubacterium, Gluconeogenesis- III, Glycolysis- VI- Metazoan, Flavin- Biosynthesis- Bacteria, Serine- Glycine- Alistipes- CAG67, Thiamin- Diphosphate- Bacteroides, Thiamin- Diphosphate, Thiamin- Diphosphate- Eukaryotes- Bacteroides, Thiamin- Diphosphate- Eukaryotes- Blautia, Tryptophan, UDP- Acetyl- Glucosamine- Ruminococcus, Valine- Ruminococcus, Valine- Unclassified.($\mathsection$)\\
    \end{tabular}
    
\end{table}

Table \ref{tab:var-compare} shows the names of pathways selected by each method for IBD vs. non-IBD, and CD vs. UC vs. non-IBD. Same microbial pathways selected in both problems are in red and vitamin related pathways are highlighted in yellow. PLDA selects all 200 pathways in some of the time points. FPCA+PLDA selects 0 pathways in IBD vs. non-IBD, and all pathways in CD vs. UC vs. non-IBD. Although more pathways are selected, the classification results are not improved compared to MFLDA-I. Thus, our method performs better in classification and simultaneously selects microbial pathways that discriminate between groups.


\end{document}